\documentclass[twocolumn,showpacs,aps,prd,nobibnotes,nofootinbib,floatfix]{revtex4-1}

\usepackage{amsmath}
\usepackage{graphicx}
\usepackage[usenames]{color}
\usepackage[normalem]{ulem}
\usepackage{textcomp}
\usepackage{gensymb}
\usepackage{xspace}
\usepackage{verbatim}

\usepackage{soul} 

\usepackage[colorlinks=true]{hyperref}
\hypersetup{
  citecolor  = blue
}

\newcommand{\beq}{\begin{equation}}
\newcommand{\eeq}{\end{equation}}
\newcommand{\MTOT}{M}
\newcommand{\MSun}{M_{\odot}}
\newcommand{\Mdot}{\dot{M}}
\newcommand{\LEM}{L_{\rm Poynt}}
\newcommand{\effEM}{\eta_{\rm EM}}
\newcommand{\LEMsteady}{L_{\rm Poynt, steady}}
\newcommand{\umagofluid}{\sigma}
\newcommand{\UNITg}{{\rm g}}
\newcommand{\UNITs}{{\rm s}}

\newcommand{\UNITcm}{{\rm cm}}
\newcommand{\UNITkm}{{\rm km}}
\newcommand{\UNITK}{{\rm K}}
\newcommand{\UNITerg}{{\rm erg}}
\newcommand{\thB}{\theta_{\rm B}}

\newcommand{\rhothir}{{\rho_{\textnormal{-}13}}}

\newcommand{\myRef}[1]{{Eq.~\ref{#1}}}
\newcommand{\myEqRef}[1]{{(Eq.~\ref{#1})}}

\newcommand{\valf}{v_{\rm Alf}}

\newcommand{\ETK}{{\texttt{EinsteinToolkit}}\xspace}
\newcommand{\Carpet}{{\texttt{Carpet}}\xspace}
\newcommand{\IGM}{{\texttt{IllinoisGRMHD}}\xspace}
\newcommand{\Pan}{{\texttt{Pandurata}}\xspace}
\newcommand{\NRPy}{{\texttt{NRPy+}}\xspace}

\definecolor{amethyst}{rgb}{0.6, 0.4, 0.8}

\definecolor{orange}{rgb}{1.0, 0.65, 0.0}

\begin{document}
\date{\today}
\title{Electromagnetic Emission from a Binary Black Hole Merger Remnant in Plasma: Field Alignment and Plasma Temperature}
\author{Bernard J. Kelly}
\affiliation{Center for Space Sciences and Technology, University of Maryland Baltimore County, 1000 Hilltop Circle Baltimore, MD 21250, USA}
\affiliation{Gravitational Astrophysics Laboratory, NASA Goddard Space Flight Center, Greenbelt, MD 20771, USA}
\affiliation{Center for Research and Exploration in Space Science and Technology, NASA Goddard Space Flight Center, Greenbelt, MD 20771, USA}
\author{Zachariah B. Etienne}
\affiliation{Department of Physics and Astronomy, West Virginia University, Morgantown, WV 26506, USA}
\author{Jacob Golomb}
\affiliation{Department of Astronomy, University of Maryland, College Park, MD 20742, USA}
\affiliation{Division of Physics, Mathematics and Astronomy, California Institute of Technology, 1200 East California Boulevard
Pasadena, CA 91125, USA}
\author{Jeremy D. Schnittman, John G. Baker, Scott C. Noble}
\affiliation{Gravitational Astrophysics Laboratory, NASA Goddard Space Flight Center, Greenbelt, MD 20771, USA}
\author{Geoffrey Ryan}
\affiliation{Department of Astronomy, University of Maryland, College Park, MD 20742, USA}
\affiliation{Astroparticle Physics Laboratory, NASA Goddard Space Flight Center, Greenbelt, MD 20771, USA}

\begin{abstract}
Comparable-mass black-hole mergers generically result in moderate to highly spinning holes, whose spacetime curvature will significantly affect nearby matter in observable ways.
We investigate how the moderate spin of a post-merger Kerr black hole immersed in a plasma with initially uniform density and uniform magnetic field affects potentially observable accretion rates and energy fluxes.
Varying the initial specific internal energy of the plasma over two decades, we find very little change in steady-state mass accretion rate or Poynting luminosity, except at the lowest internal energies, where fluxes do not exhibit steady-state behavior during the simulation timescale.
Fixing the internal energy and varying the initial fixed magnetic-field amplitude and orientation, we find that the steady-state Poynting luminosity depends strongly on the initial field angle with respect to the black hole spin axis, while the matter accretion rate is more stable until the field angle exceeds $\sim 45\degree$.
The proto-jet formed along the black hole spin-axis conforms to a thin, elongated cylinder near the hole, while aligning with the asymptotic magnetic field at large distances. 
\end{abstract}

\maketitle

\section{Introduction}
\label{sec:intro}

Black holes are the unique end-point of massive stellar evolution in our current understanding of stellar astrophysics, informed by Einstein's general relativity (GR).
They are also the inevitable result of the merger of high-mass neutron stars, as well as of black holes of all masses, including the supermassive ones believed to reside at the center of most galaxies.
Most supermassive black holes are expected to have significant spin through accretion \cite{Gammie:2003qi};
even during the course of the merger of initially nonspinning holes, enough orbital angular momentum is retained to produce a final hole with a dimensionless spin of $\sim 0.69$.
This spin angular momentum produces an azimuthal distortion of the surrounding nearby spacetime (``frame-dragging''), acting to concentrate magnetic fields and potentially produce strong steady-state electromagnetic characteristics.
Spinning supermassive black holes power active galactic nuclei (AGN) \cite{Urry:1995mg,Netzer:2015jna}, with a radio jet likely powered by the black hole's spin, mediated by polodial magnetic field lines pinned to a surrounding accretion disk\cite{Blandford:1977ds}.

We are particularly interested in the scenario of black holes newly formed after merger in a potentially complicated gas-rich environment.
Here we expect a transition from whatever was happening through the merger toward a new quasi-steady state centered on the newly formed black hole.
Generally the black hole may form in environment characterized by a larger-scale magnetic field, perhaps anchored in the poloidal component of an accretion disk that had surrounded the premerger binary.
For some configurations of spinning black hole mergers, the final black hole may be misaligned from the core axis of the broader accretion disk and its poloidal field.
It can be particularly valuable to understand the general physics driving jet formation in this kind of environment, which may be robust against detailed variations in the turbulent local environment of the black hole at the point of merger.
Toward this we consider here a simplified scenario involving a black hole in an asymptotically uniform magnetic field within a structureless uniform distribution of plasma.
Though we are primarily interested in this as a generic model for an accreting binary just after merger, we can also recognize this as a generalization of spherical Bondi accretion with a magnetic field and a spinning black hole.

In the context of the post-merger scenario, we build on \cite{Kelly:2017xck}, hereafter referred to as Paper I, where we used the tools of numerical relativity and ideal general-relativistic magnetohydrodynamics (GRMHD) to investigate how the merger of a comparable-mass black hole binary affect a surrounding plasma.
While other studies explore circumbinary accretion disks \cite{Gold:2013zma,Gold:2014dta,Farris:2012ux,Farris:2014zjo,Shi:2015cjt}, Paper I took a deliberately simplistic approach to initial conditions, in order to elucidate the effects of the merger with minimal assumptions about likely matter configurations.
In particular, we explored how an equal-mass black hole binary merger affects an initially uniform density plasma with uniform magnetic field parallel to the binary's orbital angular momentum vector. A parameter survey was performed varying the initial plasma $\beta^{-1}$ parameter, to measure the system's dependence on the relative strength of the magnetic field.

We found that the time-development of Poynting luminosity, which may drive jet-like emissions, is relatively insensitive to aspects of the initial configuration. In particular, over a significant range of initial $\beta^{-1}$, the central magnetic field strength is effectively regulated by the gas flow to yield a Poynting luminosity of $10^{45}-10^{46} \rhothir{M_8}^2 \, \UNITerg \, \UNITs^{-1}$, with the binary black-hole mass $M$ scaled to $M_8 \equiv M/(10^8 \MSun)$ and ambient density $\rhothir \equiv \rho/(10^{-13} \, \UNITg \, \UNITcm^{-3})$.
We also calculated the direct plasma synchrotron emissions processed through geodesic ray-tracing. Despite lensing effects and dynamics, we found the observed synchrotron flux varies little leading up to merger.

Here we extend the results of Paper I, paying special focus to the plasma dynamics near the remnant black hole.
In particular, we concentrate on the initial magnetic field angle relative to the remnant black hole's spin axis, and on the initial temperature of the plasma.

Great uncertainty still exists about the environment immediately around supermassive black holes. Even for Sgr A$^\ast$, the closest, most-studied black hole in the universe, the best estimates for plasma temperature, density, and magnetic field strength vary by orders of magnitude \cite{Wang:2013dqq,Bower:2018wsw,Bower:2019rsg,Corrales:2020jin}.
Similarly, the low-density gas around M87 also appears to be described by a radiatively inefficient accretion flow, but produces powerful radio jets on enormous galactic scales \cite{Reynolds:1996is,DiMatteo:2002hif,Dexter:2011xa,Prieto:2015efa}.
Therefore we acknowledge that the parameters used in this paper represent only a small region of the potential astrophysical parameter space, but we will show that these idealized conditions still provide valuable insight into some of the fundamental questions about the behavior of magnetized accretion flows around supermassive black holes.

We begin by describing the numerical methods we used in our simulations in Section~\ref{sec:methods}.
In Section~\ref{sec:newruns}, we provide the details of the parameter space survey, and in Section~\ref{sec:results}, we present results from all configurations considered.
Subsection~\ref{ssec:results_temp} considers how varying the initial specific internal energy (a proxy for temperature) affects bulk behavior such as mass accretion rates and Poynting luminosity.
Subsection~\ref{ssec:results_Bdir} considers how varying the initial magnetic field direction affects bulk behavior such as mass accretion rates and Poynting luminosity.
In Section ~\ref{ssec:jet_features}, we more closely investigate the nature of the ``proto-jet''
region that develops in the vicinity of rotating spacetimes, introducing several measures to help quantify the jet features. Throughout our paper, unless otherwise noted, we use geometrized units where $G=c=1$, and Greek (Latin) indices are space-time (space) indices.

\section{Methods}
\label{sec:methods}

In Paper I, the aforementioned black hole binary simulations
in an initially uniform plasma were carried out using the ``moving puncture'' formalism \cite{Baker:2005vv,Campanelli:2005dd},
with a simultaneous evolution of the space-time metric and MHD fields, using the \texttt{McLachlan} \cite{Brown:2008sb,mclachlan_web} implementation
of the BSSNOK equations \cite{Nakamura:1987zz,Shibata:1995we,Baumgarte:1998te} for the former, and the \IGM~\cite{Etienne:2015cea,Noble:2005gf} implementation of the
conservative GRMHD equations (see e.g. \cite{Font:2008fka}).

For the new simulations presented here, since we consider the post-merger end-state of the system,
it is more computationally efficient to use a fixed Kerr background with mass and spin appropriate
to the spacetime after the merger of an equal-mass, nonspinning binary
with initial ADM mass $M=1$: $M = 0.97$, $a/M = 0.69$
\cite{Pretorius:2005gq,Campanelli:2005dd,Baker:2005vv,Scheel:2008rj}.
There are still infinitely many ways to express such a spacetime as a metric; we choose the
horizon-penetrating ``Kerr-Schild'' slicing used by \cite{Gammie:2003rj,McKinney:2004ka}, as implemented by \NRPy \cite{Ruchlin:2017com,nrpy_web}.
This slicing has the advantage of placing the horizon at a fixed constant radial coordinate value, identical to that of the better-known Boyer-Lindquist slicing: $r_{\rm hor} = r_+ = M + \sqrt{M^2-a^2}$.
It has been used for \IGM evolutions of Fishbone-Moncrief initial conditions as part of the community Event Horizon Telescope comparison project \cite{Porth:2019wxk} and validated within the \NRPy infrastructure to satisfy the ADM constraints. 
While the numerical simulations of each configuration use the Einstein Toolkit's Cartesian mesh-refinement driver called \texttt{Carpet} \cite{Loffler:2011ay,etk_web}, for the purpose of post-processing data analysis with \texttt{GRMHD\_analysis}, a suite of Python-based tools~\cite{web:DAnalysisgitrepo}, we regularly interpolate MHD fields to a spherical polar grid.
Additionally, avoiding the spacetime evolution greatly reduces the computational cost of our studies.

Our primary diagnostic is again the EM (Poynting) luminosity:
\begin{equation}
\LEM \equiv \lim_{r \rightarrow \infty} \oint r^2 S^r d\Omega, \label{eq:LEM_def}
\end{equation}
where $S^r$ is the radial component of the relativistic Poynting vector, expressed in terms of the fluid four-velocity
$u^a$ and magnetic four-vector $b^a$:
\begin{equation}
S^i \equiv \alpha T^i_{{\rm EM}, 0} = \alpha \left( b^2 u^i u_0 + \frac{1}{2} b^2 g^i_{\;0} - b^i b_0 \right). \label{eq:Poynting_def}
\end{equation}

Rather than rely on the \ETK \texttt{Multipole} thorn's output of the spherical harmonic $(l,m)$ components of $\LEM$, as we did in \cite{Kelly:2017xck}, we output the 3D MHD field data onto the aforementioned spherical coordinate mesh with uniform sampling in each of the coordinate directions ($r$, $\theta$, $\phi$) and perform the analysis in post-processing.
This output procedure used first-order Lagrange polynomial interpolation, as supplied by the \ETK.

We apply the same post-processing suite to estimate the rate of
accretion of fluid into the black hole as well (while in
\cite{Kelly:2017xck} we used the \texttt{Outflow} code module in the
Einstein Toolkit). In particular, we calculate the flux of fluid
across the event horizon $S$ via:
\begin{equation}
\dot M = -\oint_{S} \sqrt\gamma \alpha D \left( v^i - \frac{\beta^i}{\alpha} \right) d\sigma_i, \label{eq:Mdot_formula}
\end{equation}
where $D \equiv \rho \alpha u^0$ is the Lorentz-weighted fluid density, and $\sigma_i$ is the ordinary (flat-space) directed surface element of the horizon.

Armed with these, the \emph{Poynting efficiency} $\effEM$ can
be computed from
\begin{equation}
\effEM \equiv \frac{\LEM}{\Mdot}. \label{eq:poynting_efficiency_MHD}
\end{equation}

\section{New \IGM Runs}
\label{sec:newruns}

Our investigations began with a canonical case, `\texttt{KS}',  of a single Kerr black hole in Kerr-Schild coordinates surrounded by plasma with uniform density and pressure, initially satisfying a polytropic equation of state $P = \kappa \rho^{\Gamma}$, with $\Gamma = 4/3$, appropriate for a radiation-pressure-dominated gas.
The plasma is threaded by a uniform-magnitude magnetic field oriented parallel to the hole's spin axis ($\hat{k}$). The initial fluid density, pressure, and magnetic field strength are the same as those used in Paper I's canonical configuration, yielding a fluid that is everywhere magnetically sub-dominant, with $\beta^{-1} = P_{\rm mag}/P_{\rm gas} = 0.025$.
The canonical configuration's pressure is dominated by the radiation $P_{\rm rad} = (a/3) T^4$, implying a temperature of $T = 2.906 \times 10^5 \UNITK$.
Working from this canonical case, we carried out two suites of simulations at moderate resolutions.

In the first suite, we kept the magnetic field oriented parallel to the spin axis, but varied the initial polytropic coefficient $\kappa$ in the uniform plasma, and thus the uniform specific internal energy $\epsilon$, and hence the gas temperature of the plasma.
These configurations are presented in Table~\ref{tab:initial_configs_temp}.

\begin{table}\footnotesize
\centering
\caption{Initial field configurations for the canonical case and temperature-varied simulations. $\valf$ is the Alfv\'en speed \myEqRef{eq:Valf_def}; $c_s$ is the fluid sound speed \myEqRef{eq:csound4o3}; the temperature $T$ is deduced assuming a radiation-dominated gas.}
\begin{tabular}{|c|c|c|c|c|c|c|c|c|}
	\hline 
	Name                                   & $\rho_0$ & $p_0$ &  $b_0$ &  $\sigma_0$ & $\epsilon_0$ & $\valf$ & $c_s$ & $T$ \\
	& & & & & & & & $\times 10^{5} \rho_{-13}^{1/4} \UNITK$\\ 
	\hline \hline
	\texttt{KS}                     &  1       &  0.2  &  0.1   &  0.005      &           0.60  & 0.0743  & 0.385 & 2.91\\ 
	\hline 
	\texttt{KS\_k2e-2}              &  "       &  0.02 &   "    &    "        &           0.06  & 0.0958  & 0.157 & 1.63\\
	\texttt{KS\_k4e-2}              &  "       &  0.04 &   "    &    "        &           0.12  & 0.0925  & 0.214 & 1.94\\ 
	\texttt{KS\_k6e-2}              &  "       &  0.06 &   "    &    "        &           0.18  & 0.0894  & 0.254 & 2.15\\ 
	\texttt{KS\_k9e-2}              &  "       &  0.09 &   "    &    "        &           0.27  & 0.0854  & 0.297 & 2.38\\ 
	\texttt{KS\_k3e-1}              &  "       &  0.3  &   "    &    "        &           0.90  & 0.0673  & 0.426 & 3.22\\ 
	\texttt{KS\_k4e-1}              &  "       &  0.4  &   "    &    "        &           1.20  & 0.0619  & 0.453 & 3.46\\ 
	\texttt{KS\_k6e-1}              &  "       &  0.6  &   "    &    "        &           1.80  & 0.0542  & 0.485 & 3.82\\ 
	\texttt{KS\_k9e-1}              &  "       &  0.9  &   "    &    "        &           2.70  & 0.0466  & 0.511 & 4.23\\ 
	\texttt{KS\_k2e0}               &  "       &  2.0  &   "    &    "        &           6.00  & 0.0333  & 0.544 & 5.17\\ 
	\hline 
\end{tabular} 
\label{tab:initial_configs_temp}
\end{table}

In the second suite, we kept the initial canonical temperature fixed, and varied the angle $\thB$ between the initial global magnetic field and the black hole spin.
The angles chosen were $15\degree$, $30\degree$, $40\degree$, $45\degree$, $50\degree$, $60\degree$, $70\degree$, $75\degree$, $80\degree$, and $90\degree$.

The basic \IGM simulations were carried out with a set of 10 nested fixed refinement levels, centered at the origin. Each level was cubical, with dimensions
$L_n \in \{ 1024.00M, 624.64M, 312.32M, 145.92M, 52.48M,\\ 26.24M, 13.12M, 6.56M, 3.80M, 2.38M \}$.
The grid spacing of the largest, coarsest grid ($L_0 = 1024M$) was $dx_0 = 20.48M$; each subsequent level of refinement used twice the resolution of the one before, with $dx_9 = M/25 = 0.04M$ for the finest level.
The entire mesh was offset by half the finest grid spacing ($dx_9/2$) in each direction, to avoid placing the curvature singularity $r=0$ on a grid point.

For analysis, the fields representing fluid density $\rho$, fluid pressure $p$, fluid three-velocity $v^i$, and magnetic field $B^i$ were interpolated onto an evenly spaced spherical-polar grid of size $r \in [0.35M,150M]$, $\theta \in [0,\pi]$, and $\phi \in [0,2\pi]$, with $N_r = 450$, $N_{\theta} = 50$, $N_{\phi} = 50$; hence $\Delta r \approx M/3$, $\Delta \theta \approx \pi/50$, $\Delta \phi \approx 2 \pi/50$.
The interpolation method used was a simple first-order Lagrange polynomial interpolation scheme, supplied by \ETK's \Carpet mesh-refinement driver \cite{carpet_web}.

\section{Results}
\label{sec:results}

\subsection{Dependence on plasma temperature}
\label{ssec:results_temp}

In Paper I, we investigated the dependence of the Poynting luminosity on initial density and
magnetic field strength while holding fixed the initial specific internal energy $\epsilon_0$.
As noted then, the luminosity should satisfy the scaling relation
\begin{equation}
\LEM(t) = \rho_0 \MTOT^2 F(t/\MTOT;\epsilon_0,\umagofluid_0) \label{eq:LEM_rhodep},
\end{equation}
where $\umagofluid_0 \equiv b_0^2/(2 \rho_0)$ is the initial ratio of magnetic to rest-mass energy
density, and $F(t/\MTOT;\epsilon_0,\umagofluid_0)$ is a dimensionless function of time.
Paper I primarily addressed the $\umagofluid_0$ dependence of $F(t)$, while leaving $\epsilon_0$
fixed.

Here we investigate the variation in infall rate and Poynting luminosity with $\epsilon_0$, which
serves as a proxy for the initial plasma temperature $T_0$. For a Gamma-law gas,
\[
p = (\Gamma-1) \rho \epsilon \Rightarrow \epsilon = \frac{p}{(\Gamma-1) \rho} = \frac{3 p}{\rho},
\]
where we have assumed $\Gamma = 4/3$. The set of configurations are presented in Table~\ref{tab:initial_configs_temp}.
In analogy to our varying of magnetic field strength in Paper I, here we vary the polytropic constant
$\kappa$ over two orders of magnitude; as a result the initial gas pressure $P_0$ and specific internal
energy $\epsilon_0$ also vary over two orders of magnitude, while the temperature varies by roughly a factor of three.

\begin{figure}[htpb]
\includegraphics[trim=0mm 0mm 0mm 0mm,clip,width=\columnwidth]{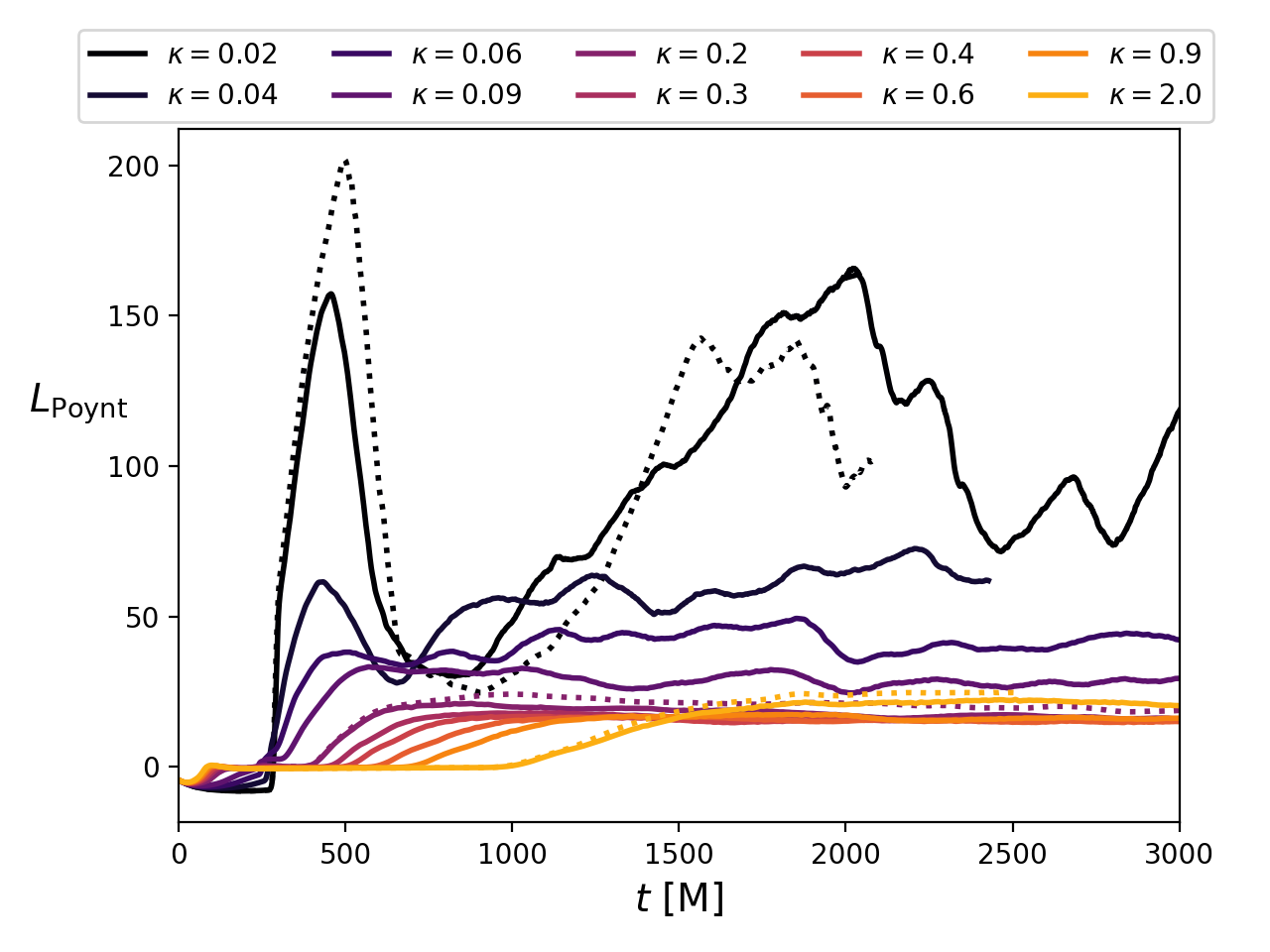}
  \caption{Poynting luminosity as a function of time for the temperature configurations listed in Table~\ref{tab:initial_configs_temp}. Dotted lines indicate data from a higher-resolution run.}
  \label{fig:LEM_temps}
\end{figure}

In Fig.~\ref{fig:LEM_temps} we show the Poynting luminosity $\LEM$ during the evolution of each of the initial temperature configurations listed in Table~\ref{tab:initial_configs_temp}.
We see that the luminosity generally takes longer to settle down with higher $\epsilon_0$, due to the lower Alfv\'en speed in these cases.
After settling, however, the late-stage luminosity shows little variation with $\epsilon_0$, except for the case of the very lowest $\epsilon_0$. This case shows extremely high luminosity, which shows no sign of settling down over the simulation time.

\begin{figure}[htpb]
\includegraphics[trim=0mm 0mm 0mm 0mm,clip,width=\columnwidth]{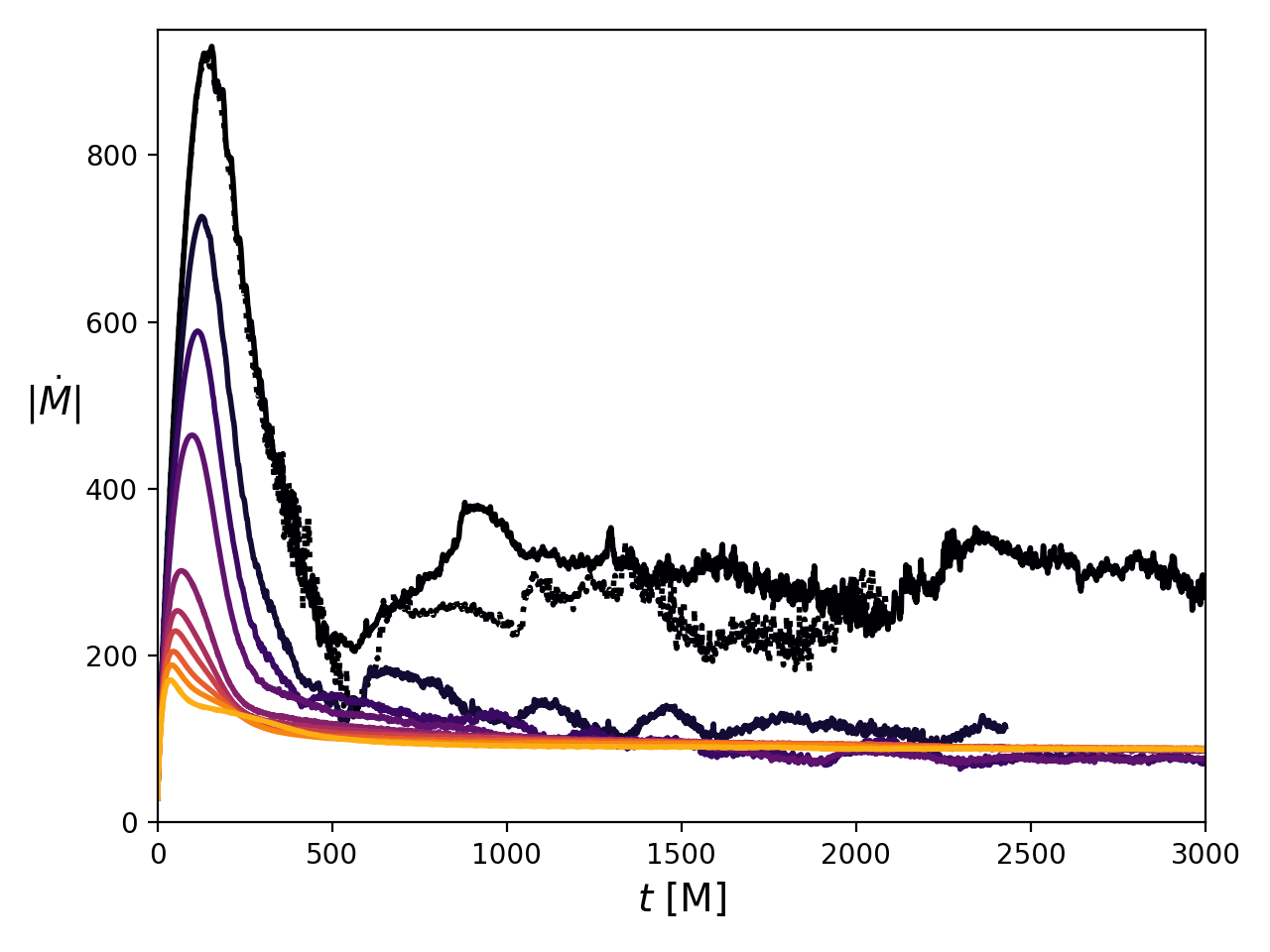}
  \caption{
  Accretion rate $|\dot{M}|$ as a function of time for the temperature configurations listed in Table~\ref{tab:initial_configs_temp}. Color labels are the same as for Fig.~\ref{fig:LEM_temps}.
  }
  \label{fig:mdot_temps}
\end{figure}

To complement the Poynting luminosity, in Fig.~\ref{fig:mdot_temps} we show the accretion rate over time of the same configurations.
Again, the lowest-temperature case, $\epsilon_0 = 0.06$, shows the least stable behavior.

\begin{figure}[htpb]
\includegraphics[trim=0mm 0mm 0mm 0mm,clip,width=\columnwidth]{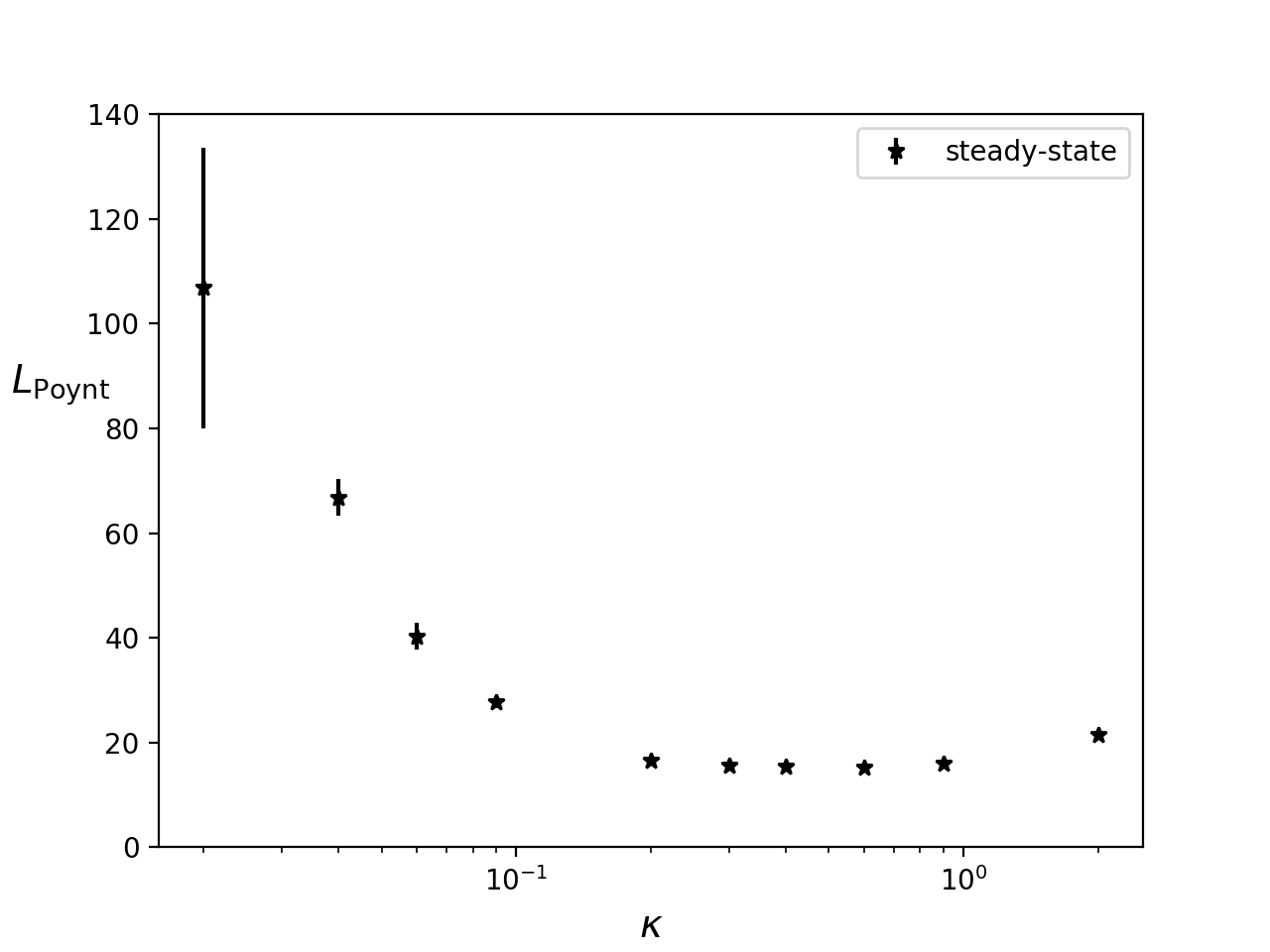}
\includegraphics[trim=0mm 0mm 0mm 0mm,clip,width=\columnwidth]{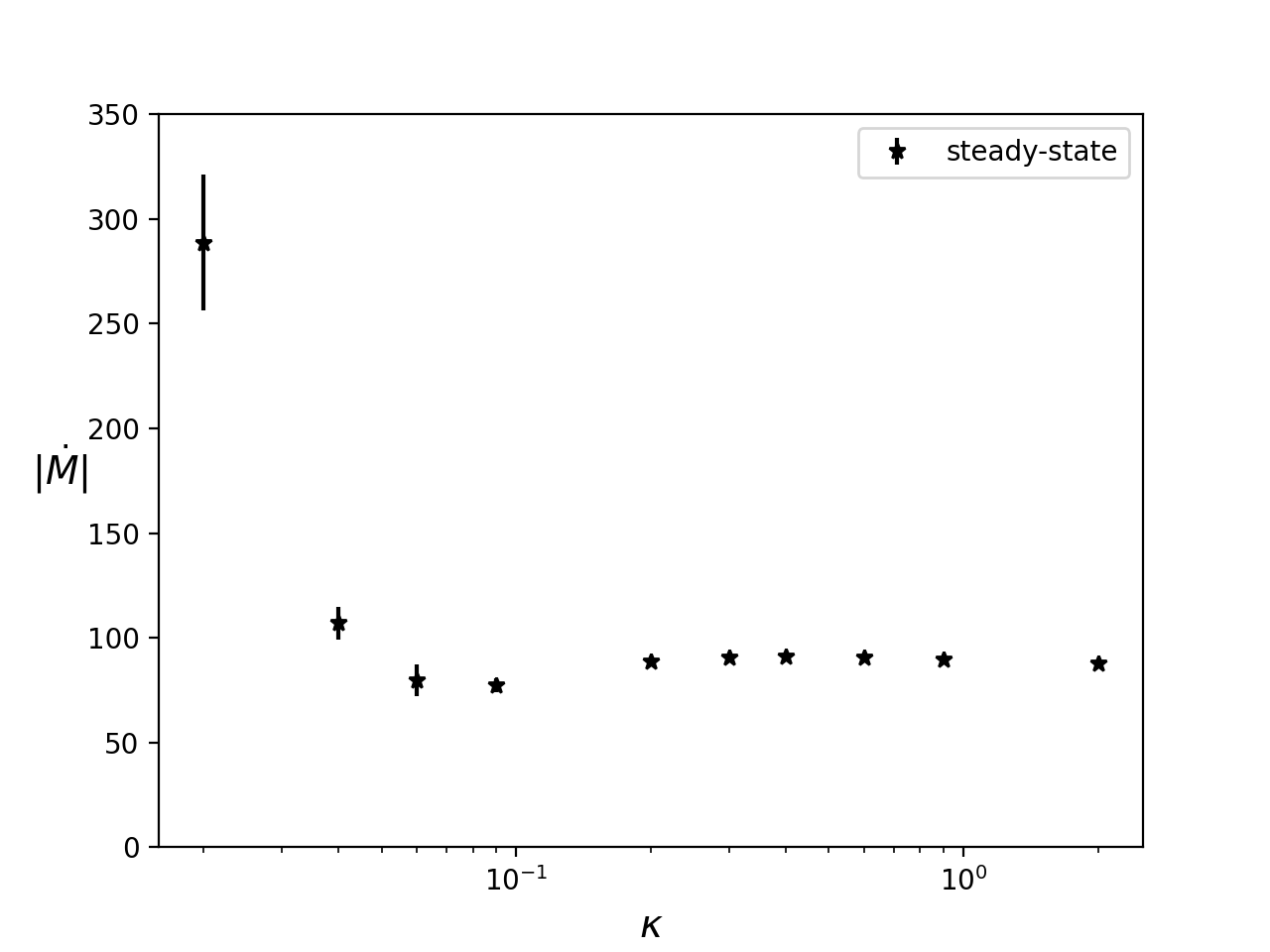}
\includegraphics[trim=0mm 0mm 0mm 0mm,clip,width=\columnwidth]{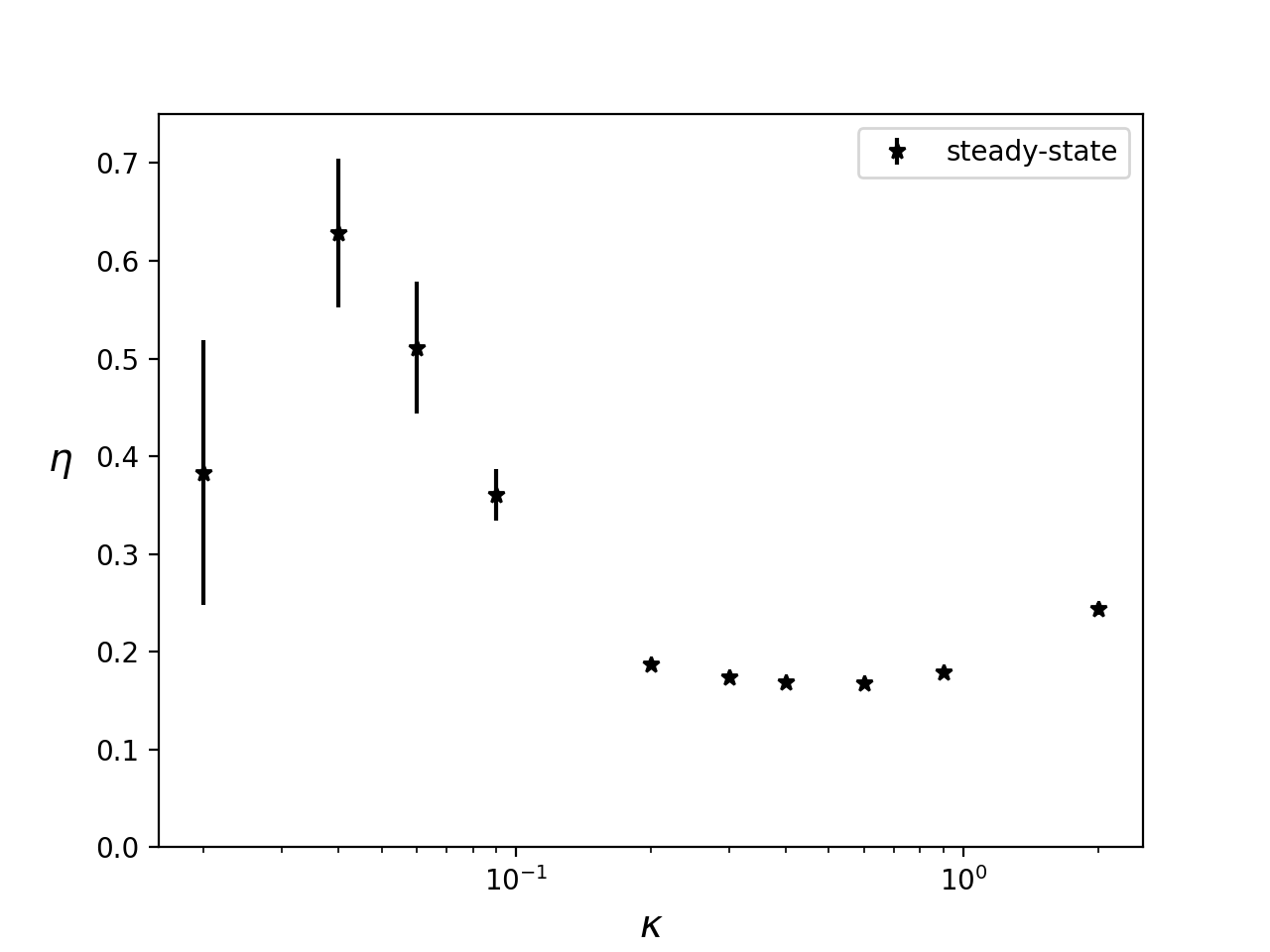}
\caption{Steady-state Poynting luminosity $\LEM$ (top panel), accretion rate $\Mdot$ (middle panel), and resulting efficiency $\effEM$ (bottom panel) for the temperature studies, as a function of the temperature proxy $\kappa$.
Plotted points are time-averages from $t = 2,000M$ onwards, with ``error bars'' given by the standard deviation over the same time interval.}
\label{fig:LEM_mdot_efficiency_steadystate_temps}
\end{figure}

The settling-down time for these configurations also depends on temperature, being later for the
higher-temperature cases.
To investigate the steady state for each configuration, we use a time-average value for each configuration, from a common starting time of $t = 2,000M$ until the end of the available data.
In Fig.~\ref{fig:LEM_mdot_efficiency_steadystate_temps}, we show the resulting Poynting luminosity $\LEM$ (top panel), accretion rate $\Mdot$ (middle panel), and resulting efficiency $\effEM$ \myEqRef{eq:poynting_efficiency_MHD} (bottom panel).
For each configuration, the ``error bars'' shown are simply the standard deviation over the time window.

It is noticeable that both the Poynting luminosity $\LEM$ and mass accretion rate $\Mdot$ are highest for the lowest values of $\kappa$, and hence fluid temperature, though subject to greater variations in time.
There also appears to be a shallow local minimum in $\LEM$ around $\kappa = 0.1$, and in $\Mdot$ around $\kappa = 0.5$; the combination of these yields a minimum in efficiency $\effEM$ around $\kappa = 0.3$, close to our canonical case.
However, as this is a shallow minimum, the efficiency is around 20\% over most of our temperature range.

\subsection{Dependence on magnetic field orientation}
\label{ssec:results_Bdir}

Here we investigate the effect of varying the angle $\thB$ between the Kerr spin vector $\vec{a}$ and the initial orientation of the uniform magnetic field $\vec{B}$.
In practice, we fix the former --- $\vec{a} = a \hat{k}$ --- and vary the latter.
However we demonstrate in Appendix~\ref{sec:robustness} that we achieve equivalent results when fixing
the field direction and varying $\vec{a}$ instead.

\begin{figure}[htpb]
\includegraphics[trim=0mm 0mm 0mm 0mm,clip,width=\columnwidth]{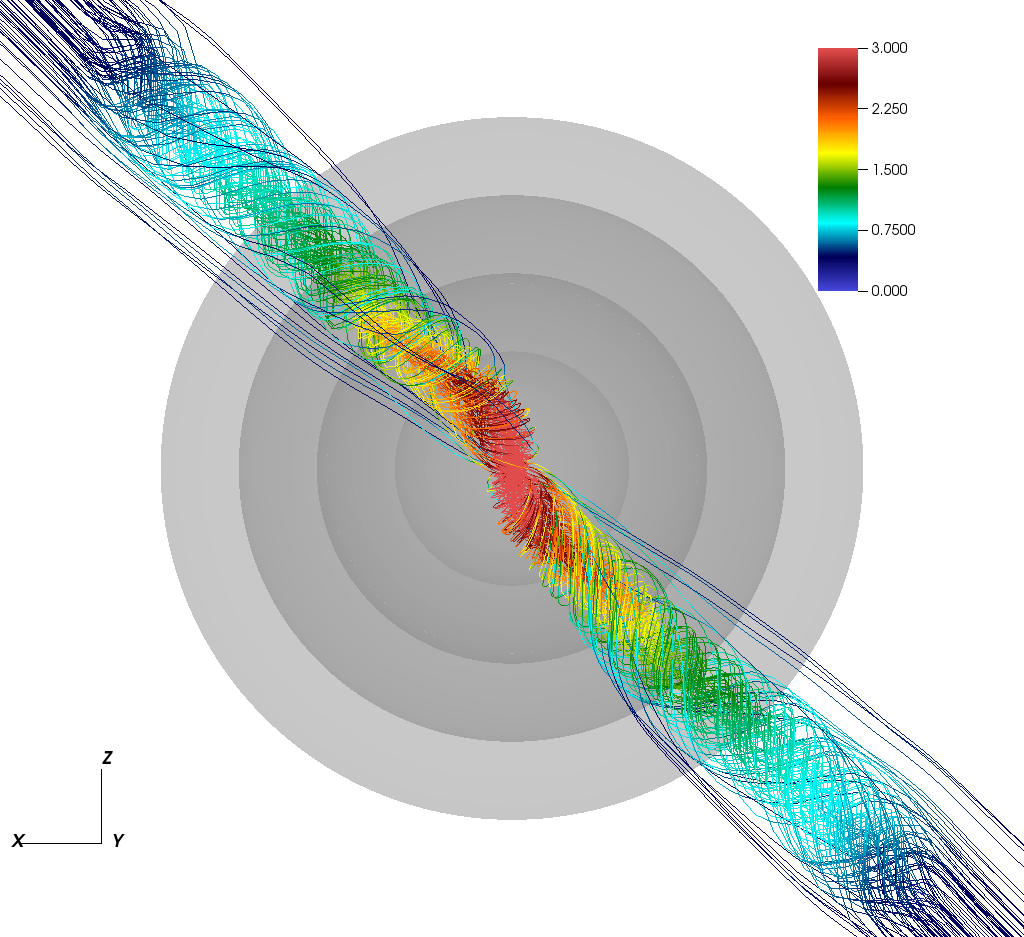}
\caption{
$B$-field stream lines in the vicinity of the BH (spinning in the $\hat{k}$ direction) at time $t \approx 2,000M$ for a magnetic field initially uniform in strength, and everywhere pointing along $\hat{i} + \hat{k}$, $45 \degree$ off the BH spin direction (configuration \texttt{KS\_B45deg}).
Grey shells indicate coordinate radii $R \in \{30M, 50M, 70M, 90M\}$.
}
\label{fig:Bxyz_B45deg_steadystate}
\end{figure}

In Fig.~\ref{fig:Bxyz_B45deg_steadystate} we show the late-time state of the magnetic field integral curves passing near the central black hole, for the \texttt{KS\_B45deg} configuration.
The black hole has not only twisted and concentrated the field, but has tilted it toward the spin axis ($z$ direction), but only out to a radius $r\lesssim 30M$.

\begin{figure}[htpb]
\includegraphics[trim=0mm 0mm 0mm 0mm,clip,width=\columnwidth]{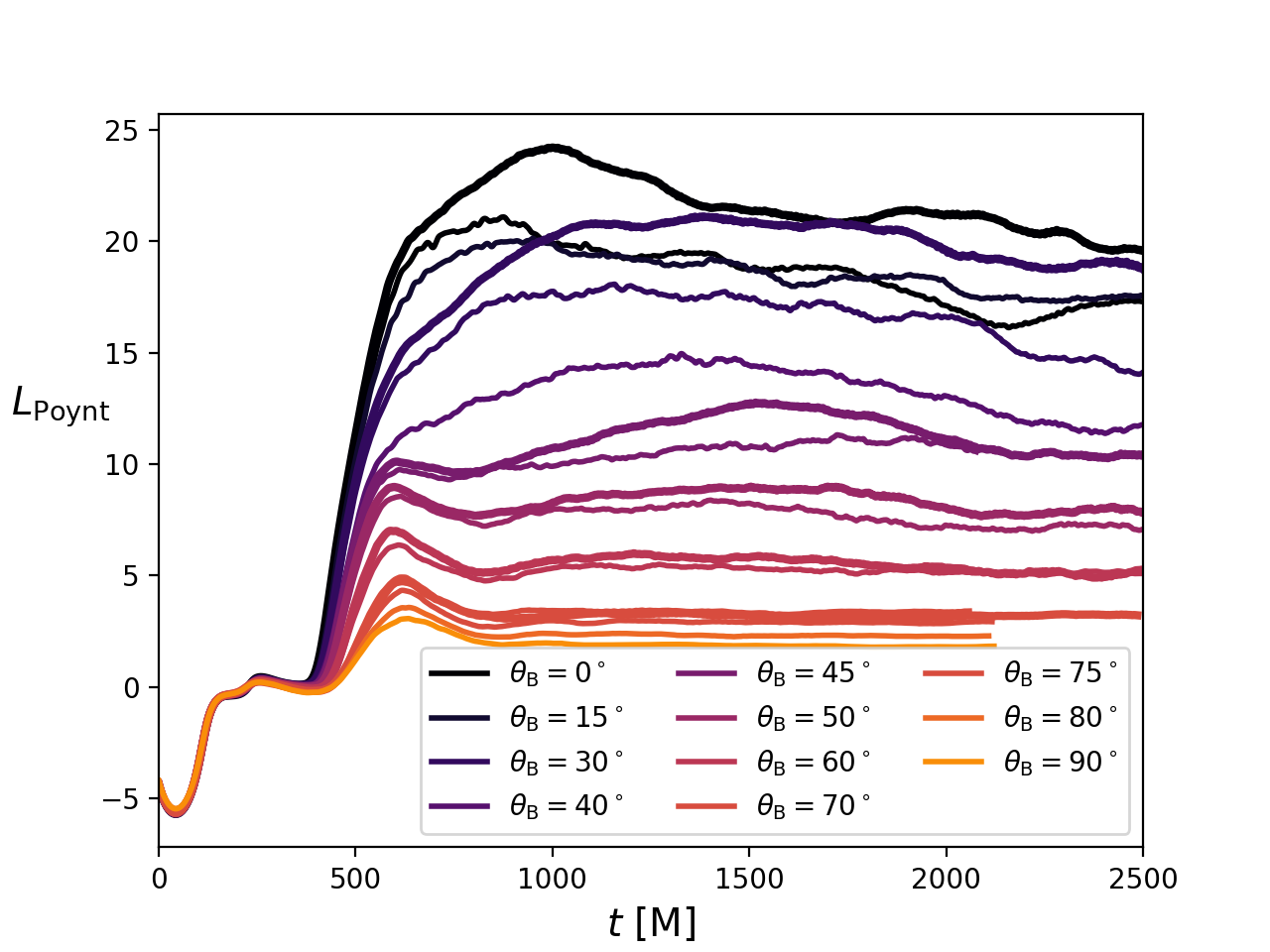}
  \caption{Poynting luminosity as a function of time for the $B$-field angle configurations.
  Thick and thin lines indicate higher and lower resolution for the same physical configuration.}
  \label{fig:LEM_Bangles}
\end{figure}

In Fig.~\ref{fig:LEM_Bangles} we show the time-development of the Poynting luminosity $\LEM$ during the evolution of each of the initial magnetic-field orientations $\thB$.
It is clear that the ``post-settling'' luminosity has a strong dependence on $\thB$.
 
\begin{figure}[htpb]
\includegraphics[trim=0mm 0mm 0mm 0mm,clip,width=\columnwidth]{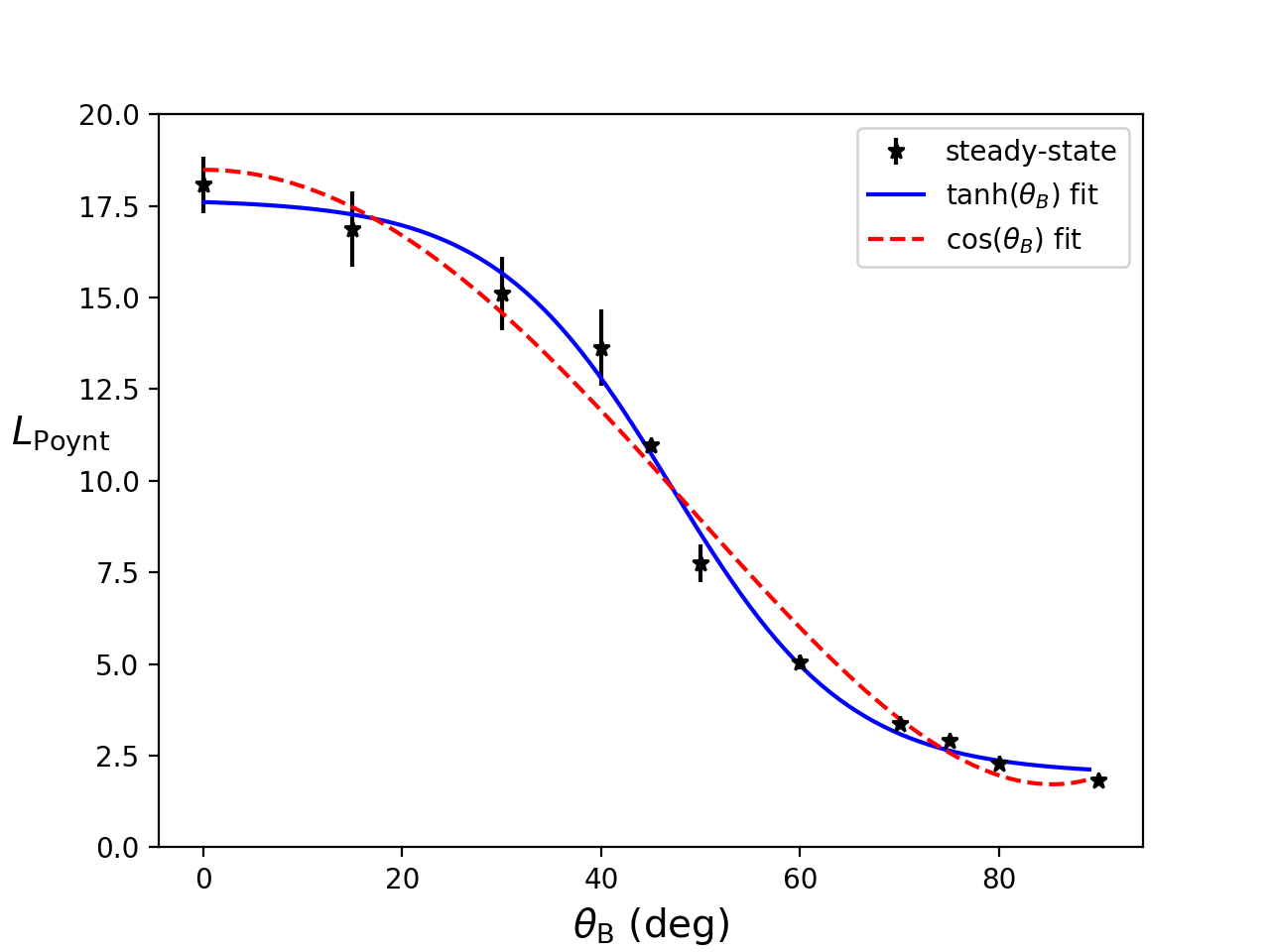}
  \caption{
  Steady-state ($t > 1,500M$) Poynting luminosity as a function of field alignment angle $\thB$. The luminosity is calculated as a ``late-time average'' value in each case --- the average value for all $t > 1,500M$.
  Error bars show the RMS deviation from the time-average values, beginning at $t = 1,500M$. The solid (blue) and dashed (red) curves are best-fit results from assuming a hyperbolic tangent or cosine-squared dependence on $\thB$, respectively.
  }
  \label{fig:LEM_Bangles_steadystate}
\end{figure}

Looking at the late-time ($t \gtrsim 1,500M$) behavior of the systems, in Fig.~\ref{fig:LEM_Bangles_steadystate} we plot $\LEM$ as a function of initial inclination angle $\thB$.
We also show a fit (dashed red line) of these $\LEM$ data points to a functional form quadratic in the cosine of $\thB$, similar to that seen by \cite{Palenzuela:2010xn} in the force-free limit. Our results seem to show a flatter behavior at low and high $\thB$, captured better by a hyperbolic tangent dependence on $\thB$ (solid blue line), but cannot rule out the $\cos^2\thB$ scaling.
It is entirely possible that the inclusion of MHD and matter (as opposed to the force-free scenario) introduces
additional physics scaling that lead to a steeper, more step-function-like behavior.

\begin{figure}[htpb]
\includegraphics[trim=0mm 0mm 0mm 0mm,clip,width=\columnwidth]{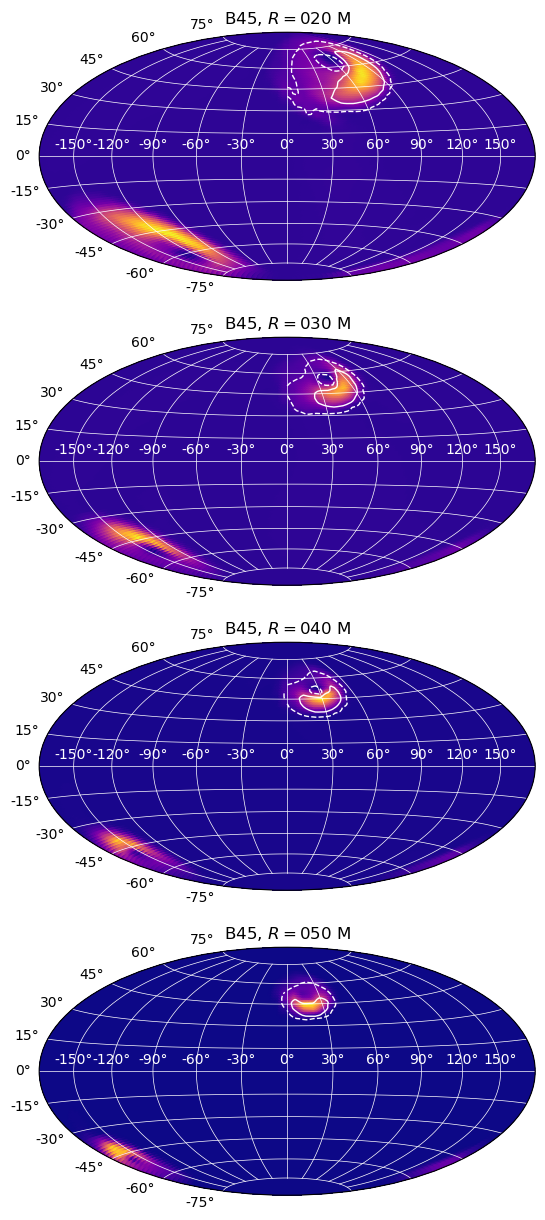}
  \caption{
  Local integral contribution to \myRef{eq:LEM_def} as a function of $(\theta,\phi)$ for extraction at $R = 20M$, $30M$, $40M$, and $50M$ for the \texttt{KS\_B45deg} configuration.
   The solid (dashed) white contours in the northern hemisphere show the regions enclosing 50\% (90\%) of the contribution to the total Poynting luminosity.
  }
  \label{fig:LEM_Pvec_2D_B45}
\end{figure}

As we can see in Fig.~\ref{fig:Bxyz_B45deg_steadystate}, even at late times, the magnetic field lines are only oriented toward the BH spin axis relatively close to the hole itself, remaining substantially along its initial direction further out.
We can try to quantify the transition region from the BH's ``sphere of influence'' by examining the Poynting luminosity over a set of extraction spheres. In Fig.~\ref{fig:LEM_Pvec_2D_B45}, we show the integrand in \myRef{eq:LEM_def} --- essentially the Poynting \emph{vector}, weighted by the local area measure --- as a function of $(\theta,\phi)$ for $R \in \{20M,30M,40M,50M\}$ for the \texttt{KS\_B45deg} configuration. We see that the angular location (i.e. ``point in the sky'') of peak contribution moves with extraction radius; we also see that the tube seems to \emph{contract} in angle.
We will attempt to quantify these observations in Sec.~\ref{ssec:jet_features}.

\begin{figure}[htpb]
\includegraphics[trim=0mm 0mm 0mm 0mm,clip,width=\columnwidth]{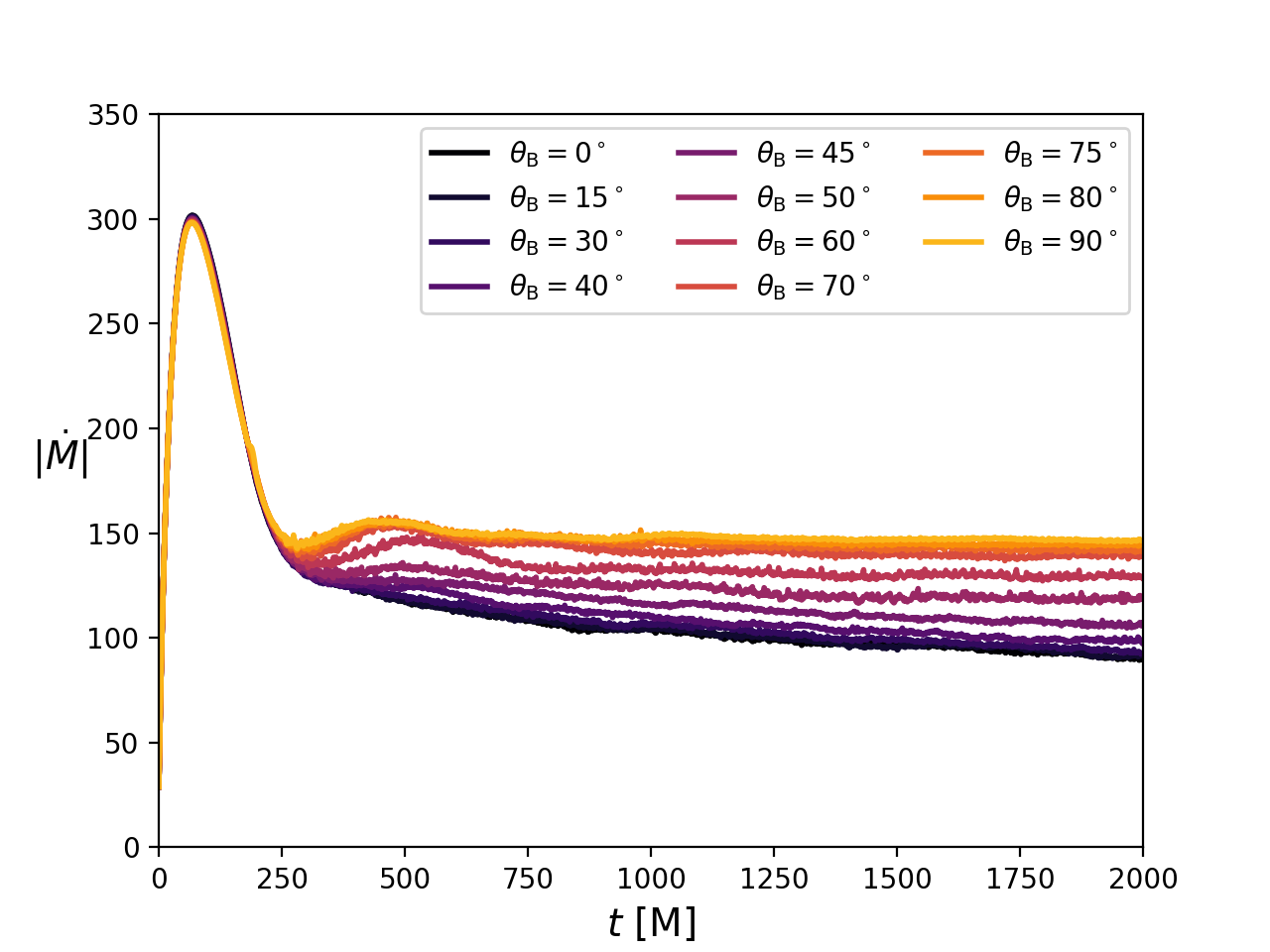}
  \caption{Accretion rate $|\dot{M}|$ as a function of time for the $B$-field angle configurations.}
  \label{fig:Mdot_Bangles}
\end{figure}

\begin{figure}[htpb]
\includegraphics[trim=0mm 0mm 0mm 0mm,clip,width=\columnwidth]{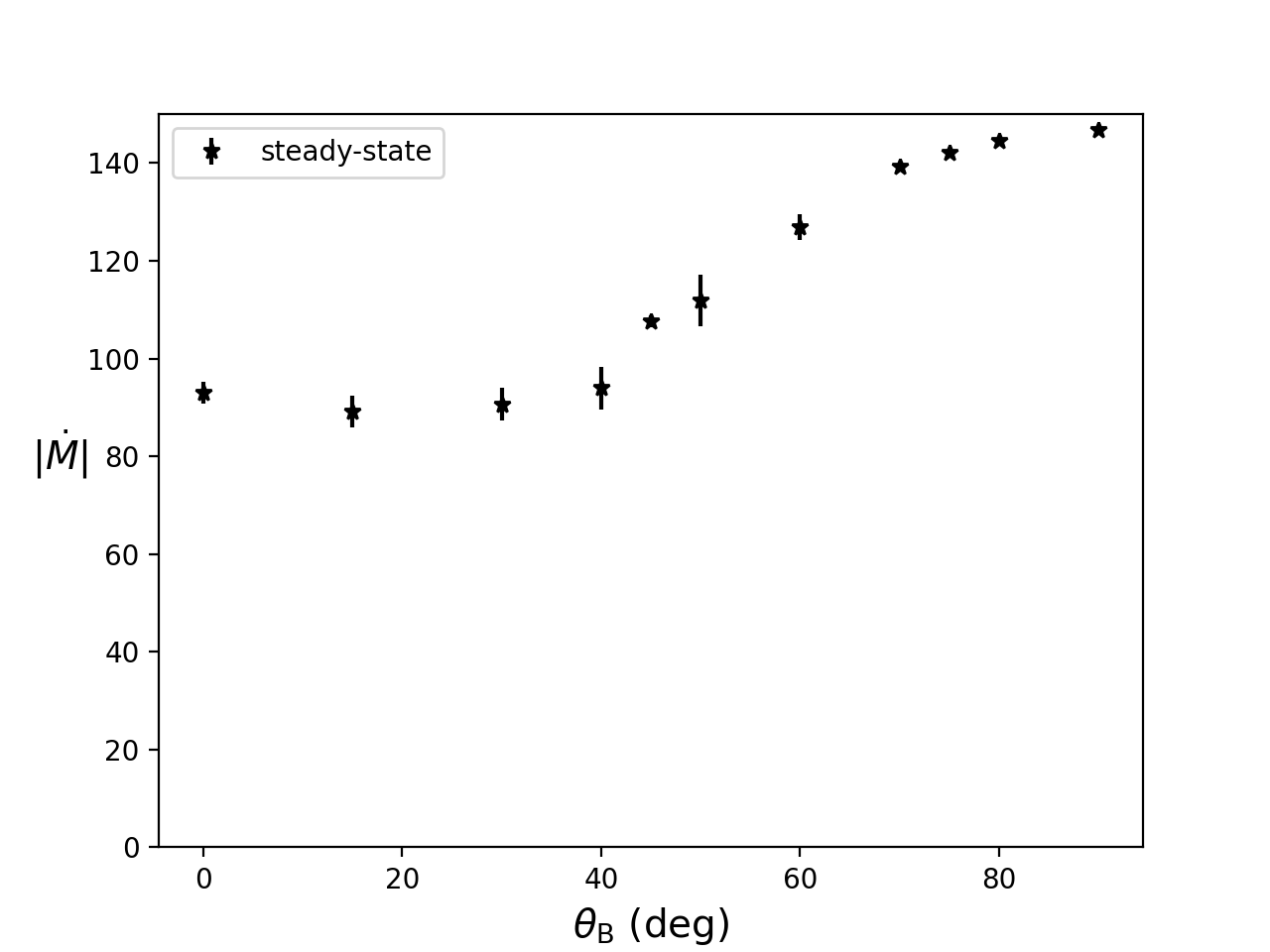}
  \caption{Steady-state ($t > 1,500M$) accretion rate as a function of field alignment angle $\theta_{\rm B}$. Error bars show the RMS deviation from the time-average, beginning at $t = 1,500M$.}
  \label{fig:Mdot_Bangles_steadystate}
\end{figure}

In Fig.~\ref{fig:Mdot_Bangles}, we show the rate of mass loss into the Kerr horizon, $\Mdot$ \myEqRef{eq:Mdot_formula} during the evolution of each of the initial magnetic-field orientations $\thB$.
Again, the accretion rates for different $\thB$ show little variation until $t \approx 300M$. Even at late times, the different configurations' $\Mdot$ deviate by only around 50\%, with the highest rates associated with the greatest deviation of the initial magnetic field angle.
As with the Poynting luminosity, we can produce a time-averaged accretion for the steady state ($t > 1,500M$) of each configuration.
This is presented in Fig.~\ref{fig:Mdot_Bangles_steadystate}. Viewed in this way, we see that the steady-state accretion rate is relatively constant
for $0\degree \leq \thB \lesssim 40\degree$, dropping off steeply for larger $\thB$.

\begin{figure}[htpb]
\includegraphics[trim=0mm 0mm 0mm 0mm,clip,width=\columnwidth]{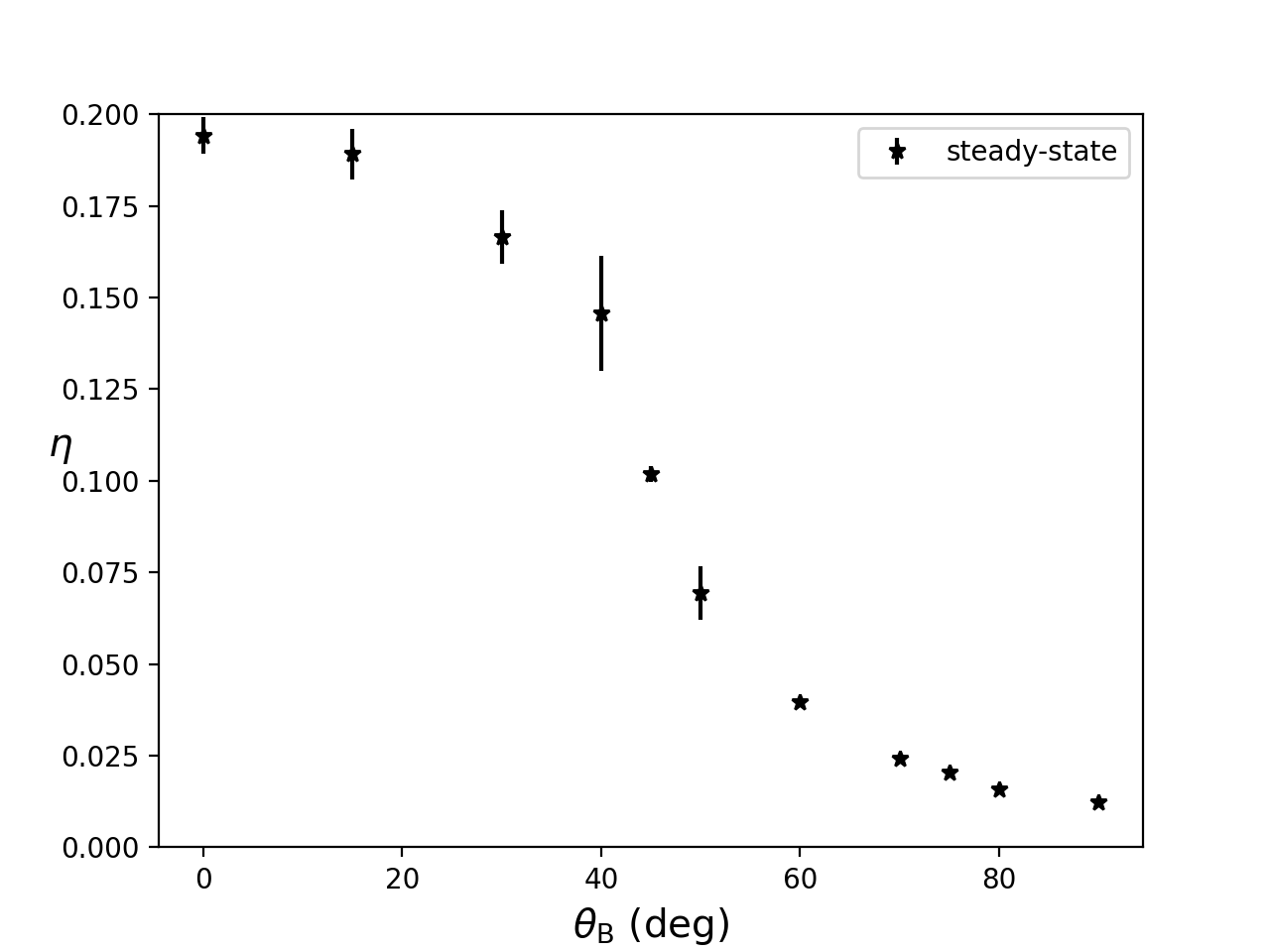}
  \caption{Steady-state ($t > 1,500M$) Poynting efficiency $\effEM$ \myEqRef{eq:poynting_efficiency_MHD} as a function of field alignment angle $\thB$.}
  \label{fig:efficiency_Bangles_steadystate}
\end{figure}

In Fig.~\ref{fig:efficiency_Bangles_steadystate}, we plot the resulting efficiency \myEqRef{eq:poynting_efficiency_MHD}. Dominated by the field orientation, it shows levels of $\sim 25\%$ for small $\thB$, dropping an order of magnitude for $\thB \gtrsim 40\degree$.

\subsection{Features of Proto-Jet}
\label{ssec:jet_features}

In studies of black-hole neutron star mergers, \cite{Paschalidis:2014qra} identify an ``incipient, magnetized jet'' as an ``unbound, collimated, mildly relativistic outflow (Lorentz factor of $\sim 1.2$), which is at least partially magnetically dominated''.
Informally, we identify a ``proto-jet'' as a magnetically dominated region showing concentrated twisting of magnetic field lines, and strong localized Poynting flux. \cite{Palenzuela:2009yr,Palenzuela:2009hx,Mosta:2009rr}
We use the term ``proto-jet'' here, because while it shows intense winding of magnetic fields in a traditional jet-like funnel region, the 
net fluid flow in this region is inward, with low Lorentz factor.
In this subsection, we attempt to clarify this definition by studying more carefully the nature of the magnetic fields and Poynting vector at late times.

\begin{figure}[htpb]
\includegraphics[trim=0mm 0mm 10mm 0mm,clip,width=\columnwidth]{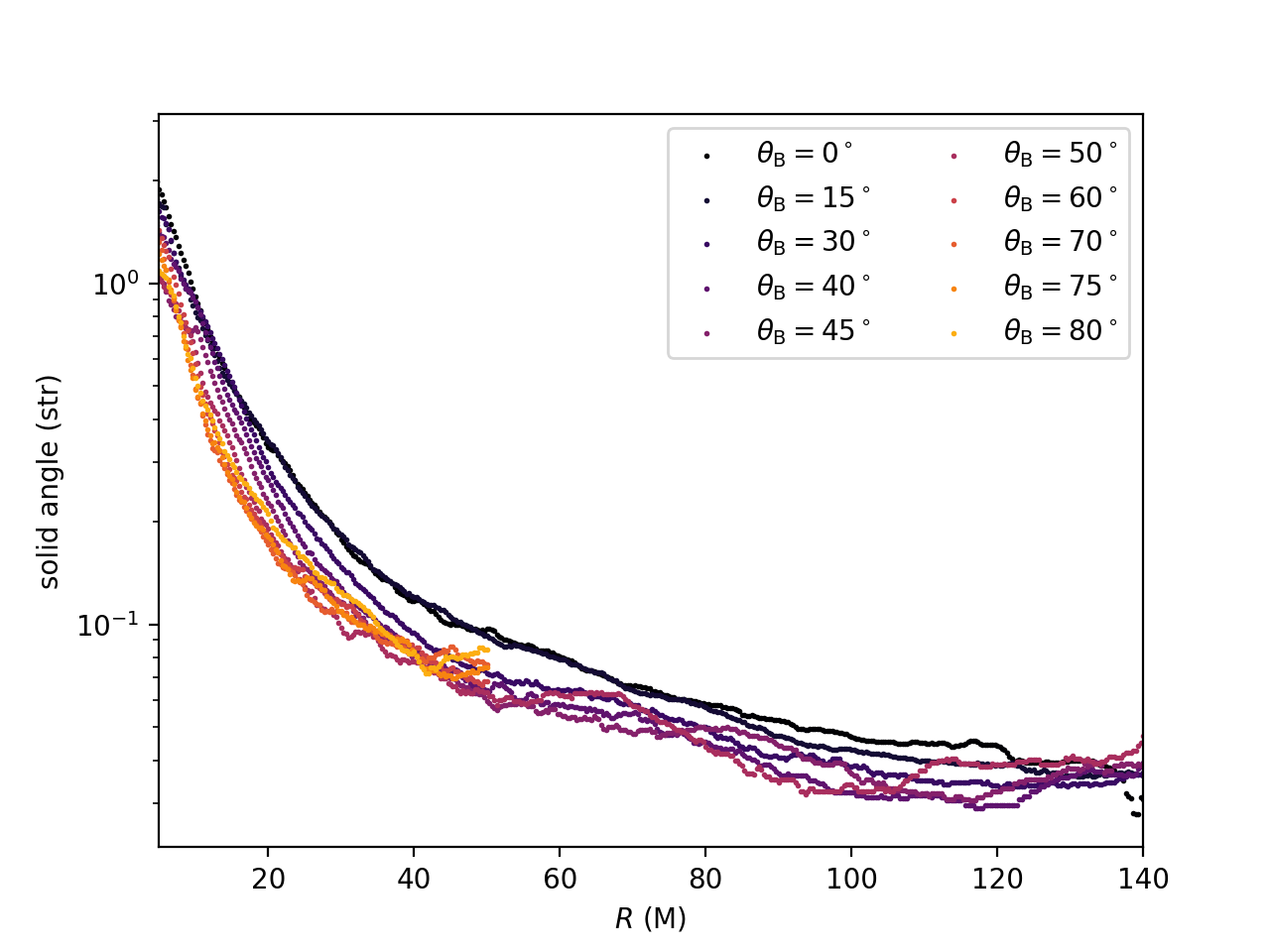}

\includegraphics[trim=0mm 0mm 10mm 0mm,clip,width=\columnwidth]{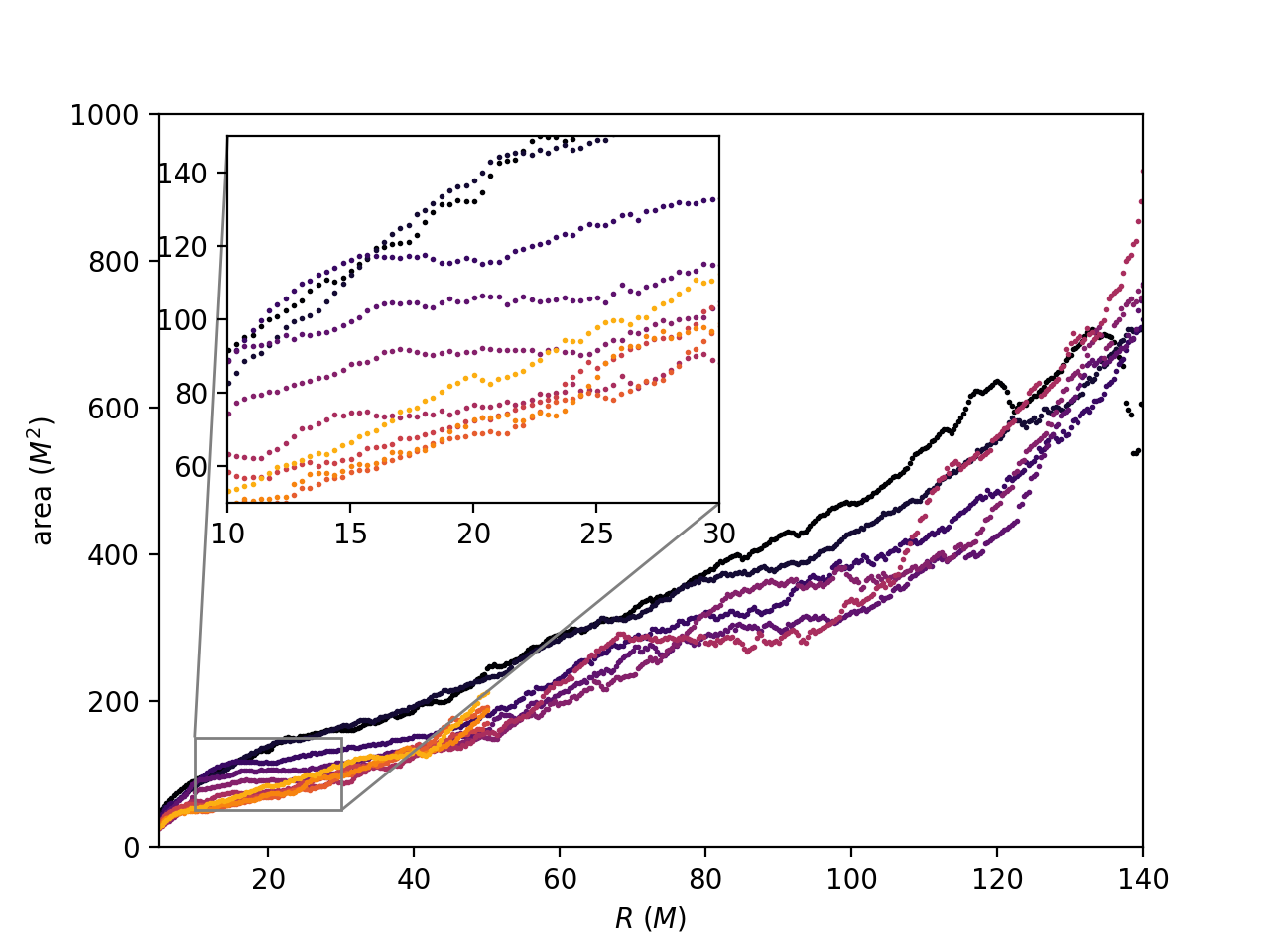}
\caption{
Upper panel: Solid angle subtended by 50\% contour of $S^r$ in the northern hemisphere at $t = 2,000M$ as a function of extraction radius $R$.
Lower panel: 50 percentile ``area'' of jet, formed by multiplying upper-panel widths by $R^2$.
The inset shows the near-leveling off of the area until $R \sim 40 M$ for intermediate configuration angles.:w
}
\label{fig:jet_widths_contours}
\end{figure}

\emph{Beam size.} To evaluate the Poynting luminosity at a radius $R$, we integrate the Poynting vector over a coordinate sphere at that radius \myEqRef{eq:LEM_def}.
Figure~\ref{fig:LEM_Pvec_2D_B45} shows the distribution of the integrand (that is, the Poynting vector weighted by the local angular Jacobian) over the sphere, with contours showing regions containing 10\%, 50\%, and 90\% of the Poynting flux.
We estimate the size of the proto-jet by calculating the solid angle subtended by the 50\% contours. At late times, we can plot this solid angle as a function of extraction radius.
In Fig.~\ref{fig:jet_widths_contours}, we show the width of the beam in the northern hemisphere as measured from the 50\% contours for each of the configurations. 
We see that the solid angular width is smaller for the larger field initial inclination angles $\thB$. Moreover, the width generally decreases with radius, especially for configurations with $\thB
\gtrsim 15\degree$ (upper panel). For extraction radii $R \in \{20M,40M\}$, the falloff in angular width is approximately $1/R^2$ -- fast enough to keep the jet's absolute cross-sectional area roughly constant, or
``pencil-like'' (lower panel)

\begin{figure}[htpb]
\includegraphics[trim=0mm 0mm 0mm 0mm,clip,width=\columnwidth]{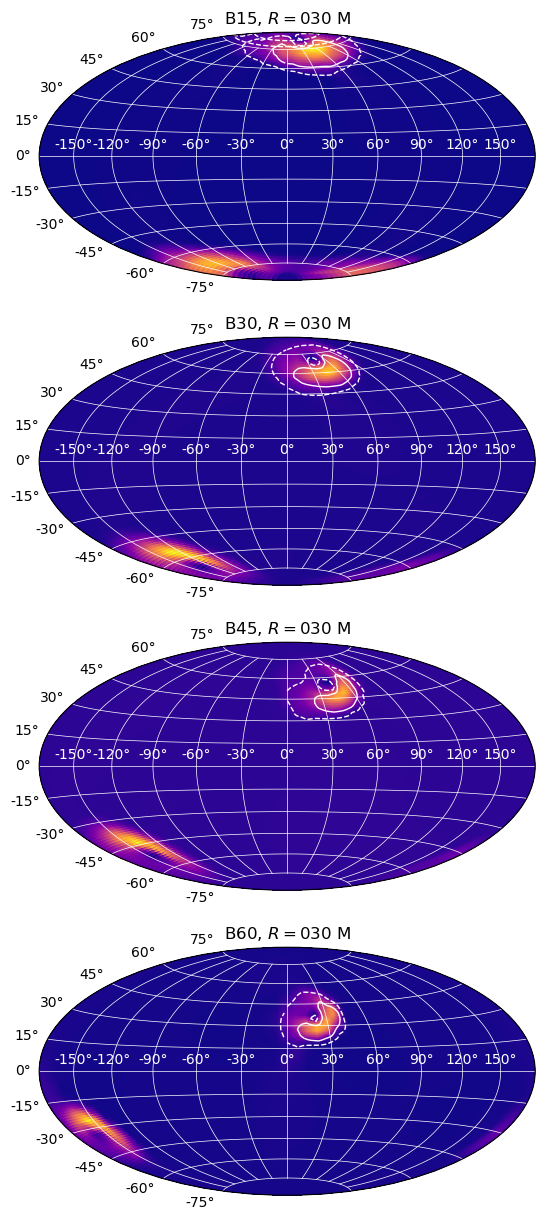}
\caption{
Local integral contribution to \myRef{eq:LEM_def} as a function of $(\theta,\phi)$ for extraction at $30M$, for magnetic-field alignments $\thB \in \{15\degree, 30\degree, 45\degree, 60\degree \}$. The solid (dashed) white contours in the northern hemisphere show the regions enclosing 50\% (90\%) of the contribution to the total Poynting luminosity.
}
\label{fig:LEM_Pvec_R30_allB}
\end{figure}

\emph{Beam shape.} As can be seen from Fig.~\ref{fig:LEM_Pvec_2D_B45}, the cross-sectional shape of the beam deviates strongly from circular when the magnetic field is misaligned with the black hole spin. We present in Fig.~\ref{fig:LEM_Pvec_R30_allB} the beam shape as represented by the 50\% contour for a range of field alignments, measured at $R=30M$.

For an aligned field, the beam cross-section is annular at all extraction radii, as the magnetic field drops to zero on the axis due to symmetry. Here we see that the beam shape becomes steadily less annular with increasing $\thB$. Simultaneously, the overall luminosity decreases, and the beam weakens, becoming harder to distinguish from the rest of the sphere. For this reason, we omit the corresponding plots for $\thB > 60\degree$.

\begin{figure}[htpb]
\includegraphics[trim=10mm 0mm 10mm 0mm,clip,width=\columnwidth]{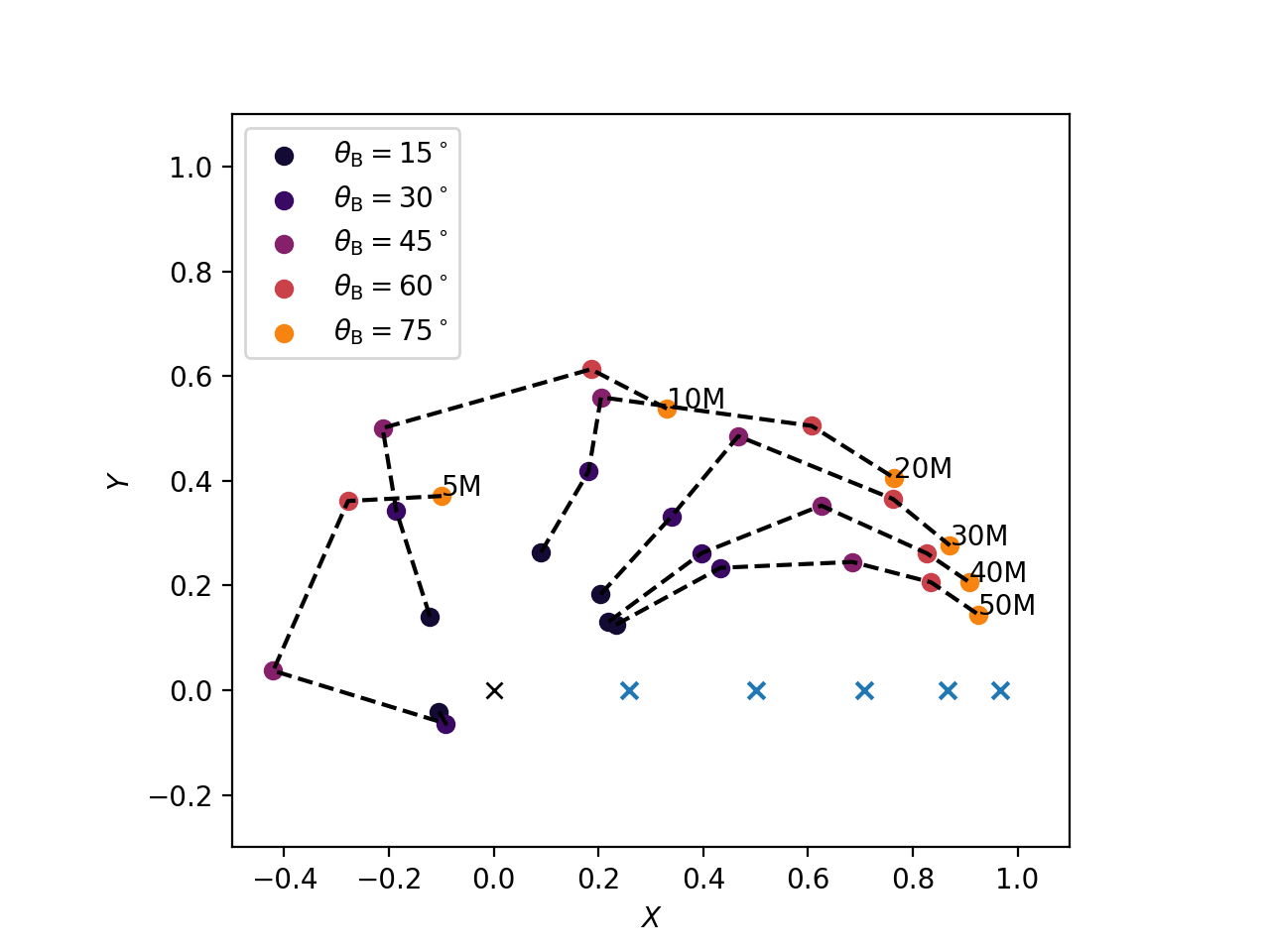}
\caption{
Pseudo-jet center positions for the B15, B30, B45, B60, and B75 configurations, in the Kerr hole's ``northern" hemisphere,
as determined by the maxima of the harmonically smoothed Poynting vector function \myEqRef{eq:Sr_ang_decomp} at $t = 2,000M$.
Each dashed line connects the positions for all configurations, determined at a certain extraction radius $R$.
The $\times$ symbols show the initial direction of the asymptotic magnetic field for each configuration.
}
\label{fig:jet_directions}
\end{figure}

\emph{Beam position.}
We present in Fig.~\ref{fig:jet_directions} the positions of the center of the
proto-jet for each configuration, showing how it varies with extraction radius. To avoid high-frequency variations, at each
extraction radius $R$, we decomposed the Poynting vector over the sphere into (real) spherical harmonics up to $\ell = 2$:
\begin{equation}
S^r_{R} (\theta,\phi) \equiv \sum_{\ell=0}^{2} S_{\ell m} Y_{\ell}^m (\theta,\phi). \label{eq:Sr_ang_decomp}
\end{equation}
The center positions are then the maxima of this smoothed functional form.
While the jet positions are properly given as a pair of angles $(\theta, \phi)$, we find it easier to display as a pair of Cartesian-like projected coordinates $X \equiv \sin\theta \cos\phi$, $Y \equiv \sin\theta \sin\phi$, so that the hole's spin direction lies unambiguously at the origin in each panel.

From the figure, we can see that all configurations have jet directions that approach the asymptotic initial magnetic field direction at large $R$ (denoted by $\times$ in the figure). As we move inward along each configuration's curve, we see twisting of the jet direction around the origin (that is, the BH spin axis). For initial inclination angles $\thB$ between $0\degree$ (i.e. parallel to the spin axis) and $\sim 60\degree$, the jet direction approaches the spin axis for small $R$. For larger $\thB$, the jet's direction stops short of the pole.

The azimuthal ($Y$-direction) offset at finite $R$ appears to be a result of frame-dragging in the background spacetime, as is the jet itself.
There is no precise transition radius where the jet direction switches from being aligned predominantly with the hole's spin to its asymptotic direction, but the transition appears to occur within $R \sim 20M$.
This is consistent with the observations of \cite{Liska:2017alm}, in their studies of jet twisting in tilted accretion tori.

\emph{Jet Strength.}
We noted at the start of this subsection that our ``proto-jet'' has not yet been demonstrated to produce ultra-relativistic particle speeds.
In particular, as in Paper I, fluid inflow in the jet region is both sub-relativistic and inward-pointing.
While analyzing the aftermath of a BHNS merger, \cite{Paschalidis:2014qra} encounter a similar situation; they point out, however, that strong magnetic dominance in the asymptotic jet region is expected to lead to much higher Lorentz factors: $\Gamma \sim b^2/2\rho$ \cite{Vlahakis:2003si}.

In our case, the peak energy ratio drops to below $\sim 5$ outside a few horizon radii, implying that actual relativistic jet conditions may not be reached for the fluid particles present.
This can be misleading, as the MHD fluid is ion-dominated, and unlikely to be the source of significant high-energy EM emission.
If a mechanism is present to seed the magnetically pressure-dominated region with electrons or electron-positron pairs, these can be expected to experience much greater accelerations, leading perhaps to jet-like electromagnetic emission.

\section{Discussion}
\label{sec:discuss}

In this paper, we have extended the work of \cite{Kelly:2017xck} (Paper I), focusing on the steady-state behavior of plasma around a post-merger Kerr black hole.
While Paper I featured merging equal-mass nonspinning black-hole binary systems, with a spacetime dynamically simulated via the moving puncture formalism, here we concentrated on the end-state of such a merger, a single spinning Kerr black hole with mass $M_{\rm final} = 0.97M$, and dimensionless spin $a/M_{\rm final} = 0.69$.
Since the spacetime is quiescent after merger, we used a fixed Kerr-Schild metric representation of the spacetime to reduce the computational load, and simplify post-simulation analysis.
Nevertheless, the initial MHD fields are fully dynamical, and our canonical MHD configuration is that of Table I of Paper I.
We have concentrated on the Poynting luminosity $\LEM$, mass accretion rate $\Mdot$, and resulting ``Poynting efficiency'' $\effEM$ of different initial plasma configurations.

First we investigated the dependence of $\LEM$ and $\Mdot$ on the specific internal energy of the plasma for fixed fluid density.
We find that higher-temperature configurations take longer to settle down to a steady state, due to the lower Alfv\'en speed in these cases.
After reaching a steady state, it is noticeable that both the Poynting luminosity $\LEM$ and mass accretion rate $\Mdot$ are highest for the lowest temperatures, though subject to greater variations in time.
There also appear to be shallow local minima in both steady-state $\LEM$ and $\Mdot$ around moderate temperatures; the combination of these yields a Poynting efficiency $\effEM$ of around 20\% over most of our temperature range, with a shallow minimum close to our canonical configuration.

Returning to the canonical configuration, we also studied the result of varying the angle $\thB$ between the asymptotic magnetic field and the Kerr hole's spin direction. We found that $\LEM$ falls swiftly with $\thB$, consistent with expectations from force-free MHD. We find the mass accretion rate is less sensitive to $\thB$ until around $40\degree$, dropping steeply thereafter.

We note here that the $\cos^2\thB$ dependence noted for Force-Free MHD \cite{Palenzuela:2010xn} was only confirmed in the low-spin limit ($a/M = 0.1$), and even then, imperfectly so.
We emphasize that the steeper behavior seen in this paper is empirical, and while we suggested a hyperbolic tangent as a ``smooth step function” that better captures the behavior seen, we do not propose that this functional form has any particular theoretical support.
We will note that (as seen in Paper I), the accompanying plasma in our Ideal GRMHD simulations induce a greater amplification of magnetic field -- and hence the Poynting luminosity -- than is seen in pure force-free simulations.
If this amplification itself is stronger for aligned-field situations, then this will cause a sharper dropoff with alignment angle than might be expected for force-free. 

Looking at the mass-accretion rate in Fig~\ref{fig:Mdot_Bangles_steadystate} further emphasizes this two-regime picture, where nearly aligned fields yield generally low mass-accretion, high Poynting luminosity, and consequently high efficiency (Fig~\ref{fig:efficiency_Bangles_steadystate}), while significantly misaligned fields show the opposite behavior.
The transition between these two regimes appears to be around $\thB = 45\degree$.

Finally, we investigated the form of the ``proto-jet'' formed by the spacetime dragging of plasma and magnetic field lines. We showed how the jet width varies with radius, and how the jet direction moves continuously from aligning with the black-hole spin axis for small $R (\lesssim 5M$) to adopting the direction of the asymptotic magnetic field for $R \gtrsim 30M$.

Having summarized the main results of our investigations, we may ask what implications these have for the astrophysical question of electromagnetic counterparts of black-hole mergers.
As noted above, we see little variation in either mass-accretion rate or Poynting luminosity over a broad range of specific internal energies around our canonical value.
We \emph{do} see a strong dependence on the angle $\thB$ between the black hole's spin and the global magnetic field, with luminosity dropping quickly as the misalignment angle increases.
Additionally, while the proto-jet is aligned with the black hole spin in the strong-gravity region (within a few Schwarzschild radii of the horizon), it soon relaxes to lie parallel to the initial asymptotic magnetic field direction.
Expectations of jet alignment are ambiguous in the absence of surrounding matter: should they align with the hole's spin or with the asymptotic magnetic field \cite{Palenzuela:2010xn,Semenov:2004ib}?
Our results indicate a transition between the two states, with the hole's spin's influence declining rapidly with distance.
Assuming that the asymptotic magnetic field is seeded by the plasma, our result here agrees qualitatively with GMRHD explorations by \cite{McKinney:2012wd} of tilted-disk simulations of highly spinning black holes, where the jet aligns with the black hole spin at $r = 4M$, but with the disk's angular momentum at $r = 40M$.

This latter observation raises the question of what we should assume for the shape, strength, and orientation of the external magnetic field.
This is a complicated question, beyond the scope of this paper, which we have neglected in favor of a survey over orientations, regardless of cause.

In astrophysical units, the steady-state luminosity for our canonical case is consistent with the ``peak'' luminosity of Paper I, but drops steeply as the angle $\thB$ between spin and asymptotic magnetic field increases, to about 10\% of its maximum.
The rate of drop-off in $\thB$ is consistent with, the $\cos(\thB)^2$ expectations from force-free models, or with a slightly steeper step-function.
We can combine the results of Paper I with the $\thB$ dependence seen in Fig.~\ref{fig:LEM_Bangles_steadystate} to obtain a more general expression for the steady-state Poynting luminosity at arbitrary spin inclination angle $\thB$ to the asymptotic magnetic field direction:
\begin{equation}
\LEMsteady \approx 10^{46} \rhothir \, M_8^2 \, H(\theta_B) \UNITerg \, \UNITs^{-1},
\end{equation}
where $H(\theta_B)$ is a function that captures the smooth step-like behavior observed in Fig.~\ref{fig:LEM_Bangles_steadystate}.

In performing the studies presented here, we have fulfilled some of the additional investigations outlined in the discussion of Paper I, focusing on the bulk behavior of MHD fields around the post-merger Kerr black hole.
Since the work here was performed on a post-merger stationary Kerr background spacetime, a full radiation transport treatment of the resulting MHD fields using, e.g. the \Pan code \cite{Schnittman:2013lka} could not be expected to produce novel results including an EM signature of the merger process, and we did not attempt it here.

Meanwhile, other extensions of Paper I are being carried out in parallel to this work, looking at the effect of significant spins on the \emph{pre-merger} black holes \cite{Cattorini:2021elw}.
With these in place, we anticipate turning our attention to rotationally supported matter distributions and magnetic field configurations, and to more realistic inclusion of radiation transport with the fully dynamical merging binary metric, using new developments in \Pan.

During review of this paper, we became aware of complementary angular studies being carried out using the \texttt{Athena++} code~\cite{Ressler:2021jjr}, using a higher central black-hole spin and a $\Gamma = 5/3$ nonrelativistic plasma, at a lower temperature than our canonical case.
While broad conclusions from that work are consistent with ours here, the different conditions do give rise to some significant differences, including a maximum jet power at intermediate angles $\thB$, rather than the monotonic decline we observe with increasing $\thB$.
These differences suggest a richer parameter space still waits to be explored in future work.

\acknowledgments

Support for this research was provided by NASA’s Astrophysics Science Division Research Program.
B.~J.~K. was supported by the NASA Goddard Center for Research and Exploration in Space Science and Technology (CRESST) II Cooperative Agreement under award number 80GSFC17M0002.
S.~C.~N. was supported in part by an appointment to the NASA Postdoctoral Program at the Goddard Space Flight Center administrated by USRA through a contract with NASA. 
Z.~B.~E. gratefully acknowledges the NSF for financial support from awards OIA-1458952, PHY-1806596, and OAC-2004311; and NASA for financial support from awards ISFM-80NSSC18K0538 and TCAN-80NSSC18K1488. 
G.~R. acknowledges the support from the University of Maryland through the Joint Space Science Institute Prize Postdoctoral Fellowship.

The new numerical simulations presented in this paper were performed in part on the Pleiades cluster at the Ames Research Center, with support provided by the NASA High-End Computing (HEC) Program.
Computational resources were also provided by West Virginia University's Spruce Knob high-performance computing cluster, funded in part by NSF EPSCoR Research Infrastructure Improvement Cooperative Agreement No. 1003907, the state of West Virginia (WVEPSCoR via the Higher Education Policy Commission), and West Virginia University.

\appendix

\section{Kerr-Schild Background}
\label{sec:KS_metric}

The form of the background Kerr metric used for the evolutions is one frequently called ``Kerr-Schild''
by accretion-disk theorists \cite{McKinney:2004ka}:
\begin{align}
ds^2 = & - \left( 1 - \frac{2 M r}{\rho^2} \right) dt^2 + \frac{4 M r}{\rho^2} dt \, dr - \frac{4 a r^2 \sin^2\theta}{\rho^2} dt \, d\phi \nonumber \\
       & + \left( 1 + \frac{2 M r}{\rho^2} \right) dr^2 - 2 a \left(1 + \frac{2 M r}{\rho^2} \right) \sin^2\theta \, dr \, d\phi \nonumber \\
       & + \rho^2 d\theta^2 + \frac{A \sin^2\theta}{\rho^2} \, d\phi^2,
\end{align}
where $\Delta \equiv r^2-(2 M r)+ a^2$, $\rho^2 \equiv r^2 + a^2 \cos^2\theta$, and $A \equiv ((r^2+a^2)^2) - a^2 \Delta \sin^2\theta$.
To be used in the \ETK, this metric must be decomposed into its ``3+1'' form -- lapse function $\alpha$, shift vector $\beta^i$, and three-metric $\gamma_{ij}$, as well as the associated extrinsic curvature $K_{ij}$ --, and transformed into a Cartesian coordinate basis.
An explicit listing of these ``3+1'' fields for the Kerr-Schild metric (still in a spherical-polar coordinate basis) can be found in the appendix of \cite{Etienne:2017jmx}.

As the radial and polar coordinates here are unchanged from that of the original Boyer-Lindquist form, the
horizon is still a coordinate sphere defined by $\Delta(r) = 0$: $r_+ = M + \sqrt{M^2 - a^2}$.

\section{Diagnostics}
\label{sec:diagnostics}

The \emph{sound speed} $c_s$ of the initial fluid configuration can be calculated as \cite{Noble:2003xx}:
\begin{equation*}
c_s^2 = \frac{\partial p}{\partial \rho} = \frac{1}{h} \left( \chi + \frac{p}{\rho^2} \kappa \right),
\end{equation*}
where $\chi \equiv \left( \frac{\partial p}{\partial \rho} \right)_{\epsilon}$,  $\kappa \equiv \left( \frac{\partial p}{\partial \epsilon} \right)_{\rho}$.

For an ideal fluid, $\chi = (\Gamma - 1) \epsilon$, $\kappa = (\Gamma - 1) \rho$, and the above simplifies to
\begin{equation}
c_s^2 = \frac{\Gamma (\Gamma-1) \epsilon}{1 + \Gamma \epsilon} = \frac{4 \epsilon}{9 + 12\epsilon}, \label{eq:csound4o3}
\end{equation}
for the $\Gamma = 4/3$ fluid we use here. For our canonical case, $\epsilon_0 = 0.6$, and $c_s \approx 0.385$.

We will also be interested in quantities that may help us predict when magneto-rotational instability (MRI) is important. To this end, we calculate the Alfv\'en speed
\begin{equation}
\valf = \sqrt{\frac{b^2}{\rho (1 + \epsilon) + p + b^2}}, \label{eq:Valf_def}
\end{equation}
where $\rho$ is the baryonic density, $\epsilon$ the specific internal energy (thus $\rho \epsilon$
is the internal energy density), $p$ is the fluid pressure, and $b^2$ is the magnetic energy density.

It may be useful to compare our results with the Bondi accretion rate, even though the latter is strictly defined for hydrodynamic fluids, and on a Schwarzschild background.
From Chapter 14 and Appendix G of \cite{Shapiro83}, we find that for a polytrope with $\Gamma < 5/3$, the Bondi accretion rate is
\begin{equation}
\dot{M}_{\rm Bondi} \approx 4 \pi \lambda_s M^2 \rho_{\infty} a_{\infty}^{-3},
\end{equation}
where $\rho_{\infty} \equiv m n_{\infty}$ is the rest-mass energy density evaluated infinitely far away from the black hole,
$a_\infty$ is the asymptotic sound speed, and the constant $\lambda_s$ is
\[
\lambda_s \equiv \left(\frac{1}{2}\right)^{\frac{\Gamma+1}{2 (\Gamma-1)}} \left( \frac{5 - 3 \Gamma}{4} \right)^{-\frac{5 - 3 \Gamma}{2 (\Gamma-1)}}
\]

For the $\Gamma = 4/3$ plasma we use in these studies, $\lambda_s = 1/\sqrt{2} \approx 0.7071$. Moreover, for our canonical
plasma configuration, $\rho_{\infty} = 1$, $p_{\infty} = \kappa \rho_{\infty}^{\Gamma} = 0.2$, and $a_{\infty} = c_s$.
Then the Bondi accretion rate is (with $M = 1$):
\[
\dot{M}_{\rm Bondi, canonical} \approx 4 \pi \frac{1}{\sqrt{2}} c_s^{-3} \approx 156. 
\]

\section{Robustness of Numerical Results}
\label{sec:robustness}

\subsection{Puncture or Fixed Kerr?}

For these investigations, we assumed a fixed Kerr background in the Kerr-Schild slicing of Sec.~\ref{sec:KS_metric}. In principle, we should allow the black-hole background to react to the influx of matter, which should increase the black hole's mass while decreasing its (dimensionless) spin, as the fluid is initially at rest in the zero-angular-momentum (ZAMO) frame. To do this we would have to used puncture-like initial data and enabled feedback in the evolutions.

However, for massive ($M \sim 10^8 \MSun$) black holes in a low-density ($\sim 10^{-13} \UNITg \, \UNITcm^{-3})$) plasma, the infalling mass and angular momentum is entirely negligible.

To justify this, we performed an evolution of our canonical system using the quasi-isotropic Kerr metric \cite{Brandt:1996si} with the same Kerr parameters ($M = 0.97$, $\chi = 0.69$). This metric can be evolved with the standard puncture formalism. With matter feedback enabled (by setting the \IGM parameter \texttt{update\_Tmunu} to true), the total matter content of the domain (and hence the MHD fluid density $\rho$) is coupled to the black hole mass $M$. With $M = 1$ in code units, we chose $\rho_0 = 10^{-6}$.
With this initial code-units density, the horizon mass of the black hole increased by $\sim 20\%$ over the course of $1000 M$ of evolution.
Dropping to $\rho_0 = 10^{-8}$ and scaling $b_0$ accordingly decreased the accretion rate by a factor of $\sim 100$, indicating that the accretion roughly scales linearly with the fluid density.

A code-units density of $\rho_0$ really means $\rho_0 M/M^3$.
The length unit appearing in the denominator is $\sim 1.5 (M/\MSun) \UNITkm$; thus the density in physical units will be $\rho_0 M/(3.375 \cdot (M/\MSun)^3) \UNITkm^{-3}$.
For our canonical total mass $M = 10^8 \MSun$, this becomes
\begin{align*}
    \rho = \rho_0 \frac{10^8 \MSun}{3.375 \cdot 10^{24}} \UNITkm^{-3} \approx 50 \rho_0 \UNITg \cdot \UNITcm^{-3}
\end{align*}
Thus our choice of $\rho_0 = 10^{-6}$ in code units is equivalent to a physical density of $\sim 5 \times 10^{-5} \UNITg \cdot \UNITcm^{-3}$.
As this is more than seven orders of magnitude greater than our assumed canonical plasma density ($10^{-13} \UNITg \, \UNITcm^{-3}$), we conclude that for all configurations considered in the main text, the black hole mass could increase by no more than one part in $10^6$, even with feedback switched on.

\subsection{Varying \texorpdfstring{$B$}{Magnetic field} or \texorpdfstring{$a$}{spin} direction?}

As mentioned in Section~\ref{ssec:results_Bdir}, we investigate the $\thB$-dependence of our results
by keeping $\vec{a} \equiv a \hat{k}$, and setting $\vec{B} = B \cos\thB$. for one representative
case, however, we instead chose the spin vector to be oriented as
$\vec{a} = a \sin \pi/4 \hat{i} + a \cos \pi/4 \hat{k}$, with $\vec{B} = B \hat{k}$.

\begin{figure}[htpb]
\includegraphics[trim=0mm 0mm 0mm 0mm,clip,width=\columnwidth]{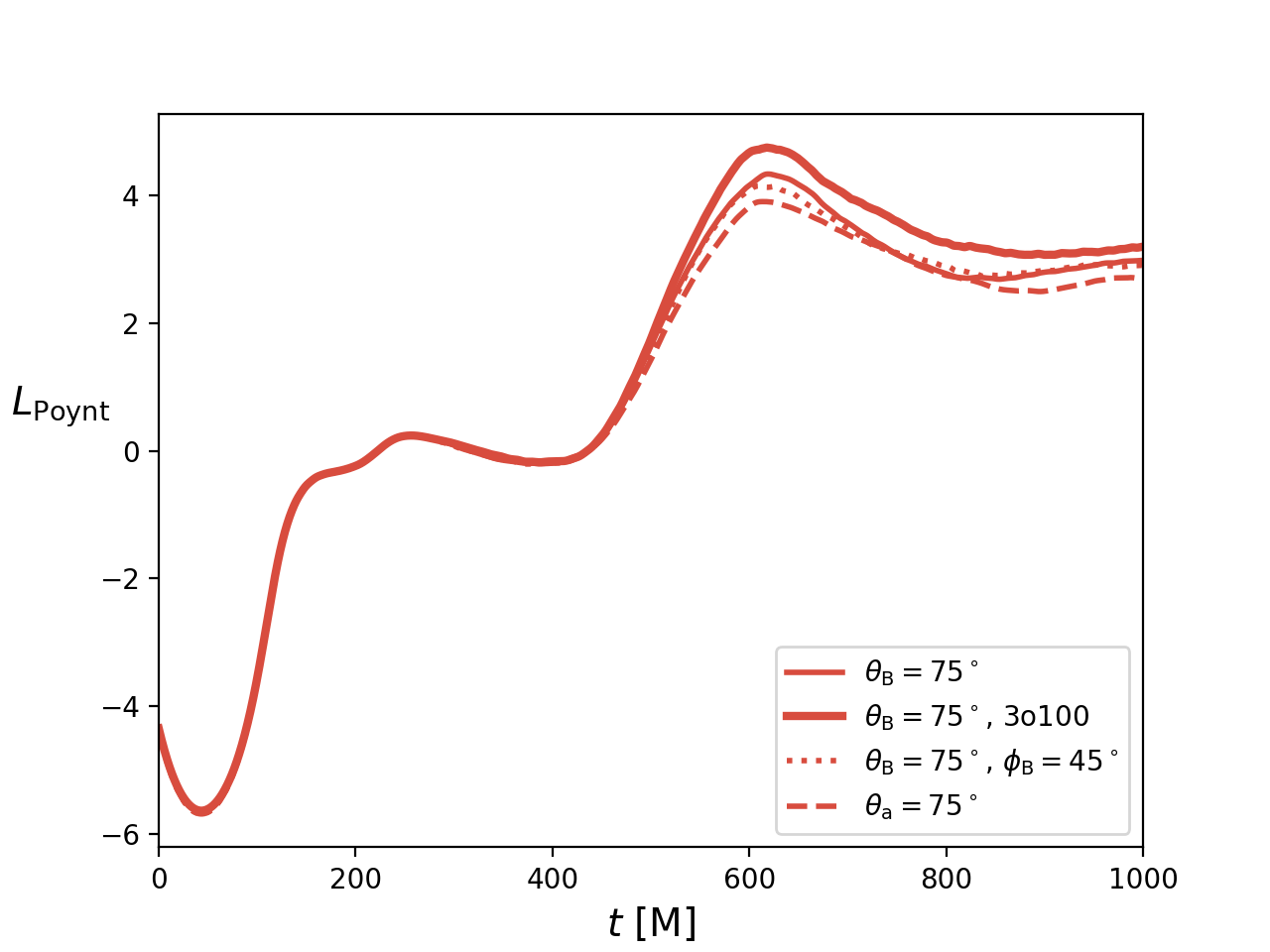}
  \caption{Poynting luminosity $\LEM$ as a function of time for $\thB = 75\degree$, tilting either the $B$-field \emph{or} the spin vector $\vec{a}$.}
  \label{fig:LEM_B75_robustness}
\end{figure}

In Fig.~\ref{fig:LEM_B75_robustness}, we show the Poynting luminosity $\LEM$ for two configurations
with $\thB = 75\degree$, with either $\vec{a}$ or $\vec{B}$ fixed along the $z$ axis. We can see that
the two curves track exactly until after the peak, where small differences begin to set in.

In this figure we also demonstrate how the luminosity changes with resolution, and with field orientation in the equatorial plane.

\end{document}